\documentclass[fleqn,usenatbib]{mnras}

\usepackage{newtxtext,newtxmath}

\usepackage[T1]{fontenc}
\usepackage{ae,aecompl}
\usepackage{csquotes}

\usepackage{graphicx}	
\usepackage{amsmath}	
\usepackage{float}
\usepackage{multicol}
\usepackage{bm}
\usepackage{subfigure}
\usepackage{relsize}
\usepackage{anyfontsize}
\usepackage[dvipsnames]{xcolor}
\usepackage{url}

\title[Fast cooling and internal heating in hyperon stars]{Fast cooling and internal heating in hyperon stars}

\author[F. Anzuini, A. Melatos, C. Dehman, D. Vigan\`o,  J. A. Pons ]{
F. Anzuini,$^{1}$\thanks{E-mail: fanzuini@student.unimelb.edu.au}
A. Melatos$^{1, 2}$, C. Dehman$^{3, 4}$, D. Vigan\`o$^{3, 4, 5}$, J. A. Pons$^{6}$
\\
$^{1}$School of Physics, University of Melbourne, Parkville, Victoria 3010, Australia\\
$^{2}$Australian Research Council Centre of Excellence for Gravitational Wave Discovery (OzGrav), University of Melbourne, \\
 \  Parkville, Victoria 3010, Australia\\
$^{3}$Institute of Space Sciences (IEEC-CSIC), Campus UAB, Carrer de Can Magrans s/n, 08193, Barcelona, Spain\\
$^{4}$Institut d'Estudis Espacials de Catalunya (IEEC), Carrer Gran Capità 2–4, 08034 Barcelona, Spain\\
$^{5}$Institute of Applied Computing \& Community Code (IAC3), University of the Balearic Islands, Palma, 07122, Spain\\
$^{6}$ Departament de Física Aplicada, Universitat d'Alacant, 03690 Alicante, Spain \\
}

\date{Accepted XXX. Received YYY; in original form ZZZ}

\pubyear{2021}

\begin{document}
\label{firstpage}
\pagerange{\pageref{firstpage}--\pageref{lastpage}}
\maketitle

\begin{abstract}
Neutron star models with maximum mass close to $2 \ M_{\odot}$ reach high central densities, which 
may activate nucleonic and hyperon direct Urca neutrino emission. To alleviate the tension between fast theoretical cooling rates and thermal luminosity observations of moderately magnetized, isolated thermally-emitting stars (with $L_{\gamma} \gtrsim 10^{31}$ erg s$^{-1}$ at $t \gtrsim 10^{5.3}$ yr), some internal heating source is required. 
The power supplied by the internal heater is estimated for both a phenomenological source in the inner crust and Joule heating due to magnetic field decay, assuming different superfluidity models and compositions of the outer stellar envelope. It is found that a thermal power of $W(t) \approx 10^{34}$ erg s$^{-1}$ allows neutron star models to match observations of moderately magnetized, isolated stars with ages $t \gtrsim 10^{5.3}$ yr. The requisite $W(t)$ can be supplied by Joule heating due to crust-confined initial magnetic configurations with (i) mixed poloidal-toroidal fields, with surface strength $B_{\textrm{dip}} = 10^{13}$ G at the pole of the dipolar poloidal component and $\sim 90$ per cent of the magnetic energy stored in the toroidal component; and (ii) poloidal-only configurations with $B_{\textrm{dip}} = 10^{14}$ G. 
\end{abstract}




\section{Introduction}
\label{sec:Introduction}
The equation of state (EoS) of neutron stars is constrained by measurements of massive objects such as PSR J$1614$--$2230$ \citep{Demorest_2010, Arzoumanian_2018}, PSR J$0348$+$0432$ \citep{Antoniadis_2013} and PSR J$0740$+$6620$ \citep{Cromartie_2020}, whose masses are close to $2 \  M_{\odot}$. Typical central densities in such heavy stars may exceed the nuclear saturation density by roughly one order of magnitude. 
However, the properties of strongly interacting matter at high densities are poorly known, and a wealth of different microphysical models have been proposed in the literature \citep{Walecka_1974,Guichon_1988, Serot_1997,Akmal_1998, Douchin_2001, Stone_2007, Lattimer_2012, Pearson_2018}. Some models predict that the ground state of baryonic matter includes concentrations of hyperons \citep{Glendenning_1985, Haensel_1994, Schaffner_2002, Stone_2007, Gusakov_2014}, which are produced when it is energetically favourable to replace high-momentum nucleons with low-momentum hyperons. 

The appearance of hyperons in neutron star cores softens the EoS, reducing below $2 \ M_{\odot}$ the maximum mass attained by stable equilibrium configurations with respect to models including only nuclear and lepton matter, in tension with the latest measurements of high-mass stars. One direct solution is that neutron stars
do not contain hyperons. Another solution, assuming the presence
of hyperons in the stellar core, is to adjust the parameters of the microphysical model including hyperons (e.g. the coupling strength among baryons and mesons) to raise the maximum mass to $\approx 2 \ M_{\odot}$ \citep{Stone_2007, Gusakov_2014, Fortin_2015, Motta_2019}. Some neutron stars with parameters adjusted in this way are predicted to cool rapidly, due to nucleonic and hyperon direct Urca emission \citep{Raduta_2018, Raduta_2019}. 

In this paper we examine the role of internal heating combined with direct Urca cooling on the thermal evolution of compact stars with and without hyperon concentrations in their core. We calculate the thermal evolution of neutron stars with masses in the range $1.1 \ M_{\odot}\leq M \leq 1.9 \ M_{\odot}$ due to both nucleonic and hyperon direct Urca emission. By way of illustration, we elect to calculate the star's profile using the $\sigma\omega\rho\phi\sigma^*$ model of nucleon-hyperon matter \citep{Gusakov_2014} and use the numerical tables corresponding to the GM1A and GM1'B parameterizations of the EoS\footnote{The numerical tables with the EoSs and the quantities needed for cooling simulations are available at \url{http://www.ioffe.ru/astro/NSG/heos/hyp.html} .}. We compare the cooling models with the observed luminosities $L_{\gamma}$ of moderately magnetized, isolated, thermally-emitting neutron stars from X-ray and optical data (e.g. \emph{Chandra} and \emph{XMM-Newton} observatories, \citep{Potekhin_2020}). The dataset comprises objects with inferred dipolar poloidal magnetic field strength at the polar surface $B_{\textrm{dip}} \lesssim 4 \times 10^{13}$ G and $L_{\gamma} \gtrsim 10^{31}$ erg s$^{-1}$ at $t \lesssim 10^{6.5}$ yr, and excludes bright objects such as magnetars and compact stars in binary systems.

A full calculation of the thermal balance in neutron stars should be conducted in two or three dimensions and include a self-consistent treatment of the stellar magnetic field. We employ the two-dimensional code developed by the Alicante group \citep{Pons_2007a, Aguilera_2008, Vigano_2012, Vigano_2013, Pons_2019, Dehman_2020, Vigano_2021} suitably modified to calculate the magneto-thermal evolution of stars hosting nuclear and hyperon matter. As a first exploratory step, we calculate the thermal balance for weakly magnetized stars and scan a wide parameter range of stellar masses with and without a phenomenological heat source, for various superfluid models. We find that if neutrons pair in the triplet channel in a small fraction of the stellar core (with maximum amplitude of the energy gap of the order $\sim 0.1$ MeV \citep{Ho_2015}), the observed $L_\gamma$ is consistent with relatively light neutron stars composed of nucleon and lepton matter ($M \leq 1.4 \ M_{\odot}$) and relatively heavy hyperon stars ($1.5 \ M_{\odot} \lesssim M \lesssim 1.6 \ M_{\odot}$). If neutrons are superfluid in a large fraction of the stellar volume (triplet pairing, with maximum energy gap amplitude $\gtrsim 0.4$ MeV), we show that both low-mass and high-mass stars require internal heating \citep{Shibazaki_1989, Umeda_1993, Kaminker_2006, Gonzalez_2010} to match $L_{\gamma}$. For stars with $B_{\textrm{dip}} \lesssim 4 \times 10^{13}$ G, we infer the parameters of a phenomenological heater to match the observed $L_{\gamma}$ in the absence of Joule heating. We then consider the case of stars with $B_{\textrm{dip}} > 4 \times 10^{13}$ G and/or strong toroidal fields at birth. Joule heating due to magnetic field decay operates efficiently in the latter regime \citep{Vigano_2013, Pons_2019, Vigano_2021}, and it is one of the candidates --- although certainly not the only one --- for the internal heater. In this paper, as a first exploratory step, we simulate Joule heating self-consistently in two-dimensions in stars with hyperon cores for a limited number of initial magnetic configurations, with the intention of scanning a wider range of low-mass and high-mass star models in future work. 

This paper is organized as follows. Section \ref{sec:Cooling_Model} reviews briefly the neutrino reactions that cool down the star. It also describes the heat diffusion and magnetic induction equations that regulate the magneto-thermal evolution of neutron stars. We then introduce the phenomenological heat source employed in this work. In Section \ref{sec:Cooling_curves} we study the thermal evolution of moderately magnetized stars with non-accreted and accreted envelopes, assuming the existence of hyperon superfluid phases with or without internal heating. Joule heating in stars born with strong magnetic fields, and its ability to supply the inferred thermal power, is calculated in Section \ref{sec:Joule}.

\section{Cooling model}
\label{sec:Cooling_Model}
We describe an idealized theoretical framework for studying the thermal and magnetic evolution of neutron stars. In this paper, for the sake of definiteness, we consider the GM1A and GM1'B EoSs \citep{Gusakov_2014}. We emphasize that there are many other valid EoS choices, e.g. the quark-meson coupling (QMC) EoS \citep{Guichon_1988, Stone_2007, Motta_2019}, and we do not seek to adjudicate between them.

Section \ref{sec:Neutrino_reactions} specifies the neutrino reactions included in the model, and Section \ref{sec:superfluidity_th} discusses the effect of superfluid phases. Section \ref{sec:magneto-thermal} introduces the heat diffusion and magnetic induction equations. In Section \ref{sec:Phenomenological_heat} we introduce a phenomenological heat source in the stellar interior, and in Section \ref{sec:theory_obs} we discuss concisely the uncertainties that affect the observations of thermally emitting, isolated neutron stars.

\subsection{Neutrino reactions}
\label{sec:Neutrino_reactions}
For the first $t \approx 10^5$ yr of a neutron star's life, typical cooling curves (i.e. the redshifted photon luminosity $L_{\gamma}$ versus stellar age) depend strongly on the dominant neutrino emission mechanisms \citep{Lattimer_1991, Prakash_1992, Yakovlev_1999, Yakovlev_2001, Page_2004, Potekhin_2015}, which drain energy out from the crust and the core. If the star is sufficiently massive, the proton and hyperon concentrations in the core may exceed (depending on the EoS) the minimum threshold required for the activation of nucleonic and hyperon direct Urca reactions in the core, leading to \textit{enhanced} cooling \citep{Lattimer_1991, Prakash_1992, Haensel_1994, Yakovlev_2001}. Stars in which direct Urca processes are inactive undergo \textit{standard} cooling, governed by the modified Urca and Cooper pair neutrino emission, and the thermal evolution of the star is slower with respect to the enhanced regime. Below we review briefly the main neutrino emission channels active in the stellar interior.

The modified Urca process \citep{Yakovlev_1995, Yakovlev_1999, Yakovlev_2001} in neutron, proton and electron ($npe$) matter is given by the reactions
\begin{eqnarray}
  && n + n \rightarrow n + p + e + \bar{\nu}_e \ ,  \quad n + p + e \rightarrow n + n +\nu_e \ ,    \ \ \ \ \  
  \label{eqn:neutron_branch}
\end{eqnarray}
\begin{eqnarray}
  && n + p \rightarrow p + p + e + \bar{\nu}_e \ , \quad p + p + e \rightarrow n + p + \nu_e \ ,  \  \ \ \ 
  \label{eqn:proton_branch}
\end{eqnarray}
which are commonly referred to as the \enquote{neutron branch} and the \enquote{proton branch} respectively. Analogous reactions hold for processes involving muons instead of electrons. We follow the literature \citep{Yakovlev_2001, Raduta_2018, Raduta_2019} and exclude the modified Urca reactions involving hyperons. 

The direct Urca emission mechanism is given by baryon $\beta$-decay and lepton capture processes \citep{Prakash_1992, Yakovlev_1999, Yakovlev_2001}
\begin{eqnarray}
  && B_1 \rightarrow B_2 + l + \bar{\nu}_l \ ,  \quad B_2 + l \rightarrow B_1 + \nu_l \ ,    \ \ \ \ \  
  \label{eqn:direct_Urca}
\end{eqnarray}
where $B_1$ and $B_2$ denote nucleons or hyperons, and $l$ denotes electrons or muons.
In neutron, proton, electron and muon ($npe\mu$) matter, these reactions are allowed only if the proton fraction (i.e. the ratio between the proton and baryon number densities) is above a critical threshold set by the requirement of momentum conservation \citep{Lattimer_1991, Yakovlev_1999}. Hyperon direct Urca processes activate close to the density threshold above which hyperons appear \citep{Prakash_1992}, and add to the nucleonic direct Urca emission for the EoS considered in this work, further accelerating the cooling rate. Typically, the direct Urca mechanism operates in the innermost regions of the star \citep{Lattimer_1991, Yakovlev_1999, Yakovlev_2001} and is more efficient than modified Urca emission \citep{Yakovlev_1999}. The cooling rate is fast for stars obtained with the GM1A or GM1'B EoSs due to the activation of both nucleonic and hyperon direct Urca. Nucleonic direct Urca is active for $M \gtrsim 1.1 \ M_{\odot}$ and $M \gtrsim 0.97 \ M_{\odot}$ for the GM1A and GM1'B EoSs respectively. $\Lambda$ hyperons appear in stars with $M \gtrsim 1.49 \ M_{\odot}$ and $M \gtrsim 1.41 \ M_{\odot}$ for the GM1A and GM1'B EoSs respectively, activating direct Urca processes involving $\Lambda$ and $p$ species. $\Xi^{-}$ hyperons appear in stars with $M \gtrsim 1.67 \ M_{\odot}$ and $M \gtrsim 1.64 \ M_{\odot}$ for the GM1A and GM1'B EoSs respectively, activating direct Urca processes involving $\Xi^{-}$ and $\Lambda$ species.

Cooper pair emission of nucleons and hyperons \citep{Flowers_1976, Yakovlev_2001, Leinson_2006, Leinson_2010, Raduta_2018, Raduta_2019} occurs via the reaction
\begin{eqnarray}
    \tilde{B} + \tilde{B} \rightarrow \nu + \bar{\nu} \ ,
    \label{eqn:Cooper_emission}
\end{eqnarray}
where $\tilde{B}$ denotes a quasiparticle in a Cooper pair and the neutrino and the antineutrino can be of any lepton family. Note that the process conserves four-momentum, because $\tilde{B}$ represents a quasiparticle. We evaluate the neutrino emission by Cooper pairs using the formulae provided in \cite{Yakovlev_1999} and \cite{Yakovlev_2001} modified according to \citet{Leinson_2006} and \citet{Page_2009} for nucleons and hyperons. We include also neutrino bremsstrahlung \citep{Yakovlev_1999}, although it is less efficient in general than Urca and Cooper pair emission.

For the crust, we match smoothly the GM1A and GM1'B EoSs with the SLy4 EoS \citep{Douchin_2001} and include pair-annihilation, plasmon decay, synchrotron, and nucleon and electron bremsstrahlung neutrino emission \citep{Yakovlev_2001}.

\subsection{Superfluidity}
\label{sec:superfluidity_th}
The presence of superfluid and superconducting phases can reduce significantly the neutrino emissivity and the heat capacity of the star \citep{Yakovlev_1999, Yakovlev_2001, Potekhin_2015}. The main contribution to both quantities comes from particles near the Fermi surface, and they are affected by the presence of an energy gap in the nucleon and hyperon dispersion relations. We assume that neutrons pair in the spin singlet channel in the crust and in the triplet channel in the core, and that superconducting protons pair in the spin singlet channel in the core \citep{Yakovlev_1999, Yakovlev_2001, Page_2004, Ho_2015}. For stars hosting hyperons, we follow \cite{Raduta_2018} and assume that $\Lambda$ and $\Xi^{-}$ particles pair in the singlet channel. The hyperon gaps are large (with typical maximum amplitude $\gtrsim 1$ MeV), because in \cite{Raduta_2018} maximum attraction is assumed between hyperon pairs. For simplicity, in the following we assume that the hyperon gaps for the GM1'B parameterization are similar to the ones obtained in \cite{Raduta_2018} for the GM1A EoS. Other interaction potentials can lead to smaller gaps (cf. \cite{Takatsuka_2001} for example). 

We quote the maximum of the density-dependent critical temperatures for the superfluid species (see Appendix \ref{sec:nucleon_superfluidity} for details). For neutron superfluidity, we use the \enquote{SFB} gap in \cite{Ho_2015}, with $T_{\rm{cn, s}} \approx 5.0 \times 10^9$ K for singlet pairing in the crust and the \enquote{c} model gap in \cite{Page_2004}, with $T_{\rm{cn, t}} =  1.0 \times 10^{10} \ \textrm{K}$ for triplet pairing in the core. For proton singlet superfluidity, we use the \enquote{CCDK} model \citep{Ho_2015}, with $T_{\textrm{cp}} \approx 6.7 \times 10^9$ K.
The maxima of the critical temperatures of $\Lambda$ and $\Xi^-$ superfluidity (singlet channel) are $T_{\textrm{c}\Lambda} \approx 7.3 \times 10^{9}$ K and $T_{\textrm{c}\Xi} \approx 2.1 \times 10^{10}$ K respectively \citep{Raduta_2018}.

In this paper we consider only superfluid phases arising from the attractive interaction among identical particles. However, as pointed out in \cite{Sedrakian_2019}, Cooper pairs made of particles of different species may form in regions of the star where the Fermi surfaces of hyperons and nucleons are close. Several works find indeed an attractive interaction among nucleons and hyperons \citep{Zhou_2005, Nemura_2009, Vidana_2018, Meoto_2020, Haidenbauer_2020}, although the interaction potentials are difficult to fit experimentally. For simplicity, we neglect the formation of superfluid phases arising from the attractive interaction between nucleons and hyperons.

We emphasize that, given the lack of experimental and observational evidence for the \enquote{correct} superfluid model, the choice of the energy gaps for baryon species introduces a certain degree of arbitrariness in our results. The energy gaps affect the cooling curves (see Section \ref{sec:Y_SF} and Appendix \ref{sec:Alternative_models}) and the inferred thermal power from an internal heat source (see Section \ref{sec:Internal_heat}) required by low- and high-mass models to match $L_{\gamma}$ observations. Uncertainties in the energy gaps of nucleon and hyperon superfluid phases are a focus of extensive research \citep{Takatsuka_2001, Kaminker_2002, Page_2004, Ho_2015, Raduta_2018, Raduta_2019}. For hyperons, there is a somewhat larger uncertainty than for nucleons, given the lack of experimental constraints on hyperon-hyperon interaction potentials \citep{Balberg_1998, Takatsuka_1999, Takatsuka_2001, Wang_2010}.

\subsection{Magnetic and thermal evolution}
\label{sec:magneto-thermal}

We introduce the magnetic induction and heat diffusion  equations that regulate the magnetic and thermal evolution of neutron stars respectively, assuming that the space-time structure is described by the Schwarzschild metric and that deviations from spherical symmetry due to rotation, strong magnetic fields and temperature are negligible \citep{Pons_2019}.

\subsubsection{Magnetic induction}

In this work we consider the crust-confined magnetic field configurations studied thoroughly elsewhere, e.g. by \cite{Vigano_2013}, \cite{Wood_2015}, \cite{Gourgouliatos_2016}, \cite{Dehman_2020}, \cite{Igoshev_2021} and \cite{DeGrandis_2021}. The magnetic field evolves according to the magnetic induction equation, which in the crust reads

\begin{eqnarray}
\frac{\partial \boldsymbol{B}}{\partial t} = -\boldsymbol{\nabla}\times \Big[ \frac{c^2}{4 \pi \sigma_e} \boldsymbol{\nabla}\times (e^{\Phi}\boldsymbol{B}) + \frac{c}{4 \pi e n_e}[\boldsymbol{\nabla} \times (e^{\Phi}\boldsymbol{B})] \times \boldsymbol{B} \Big] \ ,
\label{eqn:induction}
\end{eqnarray}
where $c$ is the speed of light, $\sigma_e$ is the temperature- and density-dependent electrical conductivity, $\Phi$ is the gravitational potential, $e$ is the elementary electric charge and $n_e$ is the number density of electrons, assumed to be the only charge carriers.
The first term in Eq. \eqref{eqn:induction} is the Ohmic dissipation term, and the second is the Hall term.
The Hall term strongly enhances the dissipation of magnetic energy over the first $t \approx 10^6$ yr of the neutron star's life with respect to the purely resistive case  (i.e. when only Ohmic dissipation is active) due to two effects: (i) generation of small-scale structures, where Ohmic dissipation is more efficient; and (ii) gradual compression of the crustal electric currents toward the crust-core interface. If a highly-resistive layer is present due to impurities or pasta phases, there is an enhancement of the dissipation rate of the magnetic field at the crust-core interface. The Joule heating rate increases, as the abundance of impurities increases. Typically, Joule heating is important in stars with poloidal fields satisfying $B_{\textrm{dip}} \gtrsim 10^{14}$ G, or in the presence of strong toroidal fields. We neglect the role of plasticity in the crust, which may alter the Hall time-scale \citep{Lander_2019}.

The magnetic induction equation is solved starting from an initial magnetic field configuration. The latter is poorly known \citep{Duncan_1992,Thompson_1993, Spruit_2008}. Recent numerical simulations of the magnetorotational instability mechanism in core-collapse supernovae suggest that tangled magnetic field configurations with a large fraction of magnetic energy stored in small-scale and multipolar structures may form \citep{Aloy_2021, Reboul-Salze_2021}. Additionally, some isolated, thermally-emitting neutron stars have anisotropic surface temperature distributions which may be compatible with multipole magnetic field configurations \citep{De_Luca_2005, Potekhin_2020}, as supported by recent \textit{NICER} observations \citep{Riley_2019, Bilous_2019}. For the sake of definiteness, we limit ourselves to two initial, crust-confined and axisymmetric configurations: one that is purely poloidal (dipole), and one that is a mixture of poloidal (dipole) and toroidal (quadrupole) components. We emphasize that neglecting core-threading magnetic field topologies is an oversimplification. However, the study of the magnetic field in the core is more complex than in the crust (due to its multifluid nature and the occurrence of superconducting phases), and there is no numerically tractable formulation of the induction equation available at present (cf. \cite{Gusakov_2020} and references therein). We postpone the study of different initial topologies to future work.

\subsubsection{Heat diffusion equation}
The thermal evolution is regulated by the heat diffusion equation

\begin{eqnarray}
    c_\textrm{V}e^{\Phi}\frac{\partial T}{\partial t} + \boldsymbol{\nabla}\cdot(e^{2\Phi} \boldsymbol{F}) = e^{2\Phi}(H + Q_{\textrm{J}} - Q_{\nu}) \ ,
    \label{eq:thermal_evolution}
\end{eqnarray}
where $c_\textrm{V}$ is the heat capacity per unit volume of nucleons, leptons and hyperons and $T$ is the internal, local temperature. The heat flux is given by $\boldsymbol{F} = - e^{-\Phi} \hat{\kappa} \cdot \boldsymbol{\nabla}(e^{\Phi}T)$, where $\hat{\kappa}$ denotes the anisotropic thermal conductivity tensor \citep{Potekhin_2001, Potekhin_2003, Vigano_2021}, $Q_{\textrm{J}}$ is the Joule heating rate per unit volume and $Q_{\nu}$ is the neutrino emissivity per unit volume. The function $H$ describes the thermal power per unit volume generated by internal heating mechanisms other than Joule heating.
 
Cooling simulations \citep{Glen_1980, Yakovlev_1999, Page_2004, Potekhin_2015,  Potekhin_2018} divide the star into an internal region (with density $\rho \gtrsim 10^{10}$ g cm$^{-3}$, which becomes isothermal in the absence of strong magnetic fields after $t \approx 10^2 \ \textrm{yr}$) and a nonisothermal outer envelope \citep{Gudmundsson_1983}, which contains negligible mass and insulates thermally the stellar surface from the interior layers. The outer envelope is characterized by strong radial gradients of density, pressure and temperature, and much shorter thermal timescales (compared with the interior). Quantitatively, the gradients and, therefore, the surface temperature, depend on the chemical composition of the envelope, which is uncertain. A common assumption is that it consists of iron-like ions. However, envelopes containing lighter chemical elements like hydrogen, helium and carbon could be produced by nuclear transformations of matter accreted on the outermost layers of the star \citep{Potekhin_1997, Potekhin_2001, Potekhin_2003, Page_2004}. For an accreted envelope, the chemical composition is a function of $M$ and the accreted mass $\Delta M$ and depends on the nuclear reactions that transform the infalling matter, forming a shell structure where each layer is dominated by one chemical element \citep{Potekhin_1997,Potekhin_2001, Potekhin_2003}. Heavier elements reduce the thermal conductivity of the outer envelope, and hence reduce the surface temperature $T_\textrm{s}$ with respect to envelopes hosting lighter elements.

The magnitude and direction of the magnetic field affect $T_\textrm{s}$ as well. Electrons conduct heat differently in perpendicular and parallel to the magnetic field lines. For $10^{11} \ \textrm{G}\lesssim B_{\textrm{dip}} \lesssim 10^{13} \ \textrm{G}$, $T_\textrm{s}$ is reduced with respect to stronger and weaker fields due to the reduction of heat propagation perpendicular to the magnetic field lines \citep{Potekhin_2001, Potekhin_2003}, resulting in increased thermal insulation by the outer envelope at the equator. For $B_{\textrm{dip}} \gtrsim 10^{13}$ G the magnetic field increases the heat transport along the magnetic field lines close to the pole, the insulation of the outer envelope is reduced, and the photon luminosity is higher. 

The heat diffusion equation is solved in the core and crust, with the exception of the outer envelope. We use the solution of the heat diffusion equation at the base of the nonisothermal outer envelope ($T_{\textrm{b}}$) to obtain $T_{\textrm{s}}$ using the analytical relation $T_{\textrm{s}}$ -- $T_{\textrm{b}}$ given in \cite{Potekhin_2015} (and adopted in \cite{ Vigano_2021}).
The $T_{\textrm{s}}$ -- $T_{\textrm{b}}$ relation fits the numerical results of hydrostatic models of the envelope. The effect of the magnetic field on the surface temperature is included for heavy-element envelopes and neglected for light-element envelopes. In the following, we place the bottom of the outer envelope at $\rho_\textrm{b} \approx 4 \times 10^{10}$ g cm$^{-3}$. We note that envelope models do not account for Joule heating, which causes further dissipation of the magnetic field in the resistive envelope (see Fig. 8 in \cite{Akgun_2018}).

\subsection{Phenomenological internal heating}
\label{sec:Phenomenological_heat}
Isolated neutron stars with $B_{\textrm{dip}} \lesssim 4 \times 10^{13}$ G may attain higher-than-expected temperatures via heating mechanisms operating in their interiors different from Joule heating. For example, they can be heated up by dissipation of rotational energy due to friction between the crust and the superfluid component \citep{Alpar_1984, Shibazaki_1989,Riper_1995, Page_2006}. Stars born with millisecond rotational periods may be heated up by rotochemical heating \citep{Reisenegger_1995, Fernandez_2005, Yanagi_2020}.

A phenomenological approach to simulate internal heating mechanisms was developed in a series of papers by \cite{Kaminker_2006}, \cite{Kaminker_2007}, \cite{Kaminker_2009}, \cite{Kaminker_2014}. In this phenomenological approach, an internal source is assumed of the form

\begin{eqnarray}
H = H_0 \  \Theta(\rho_1, \rho_2) \ \textrm{e}^{-t/\tau} \ ,
\label{eq:heat}
\end{eqnarray}
which produces a total thermal power (redshifted at infinity)

\begin{eqnarray}
W(t) = \int dV e^{2\Phi} H \ .
\label{eq:heat_integral}
\end{eqnarray}
In Eq.\eqref{eq:heat}, $H_0$ is a constant in units of erg cm$^{-3}$ s$^{-1}$, $\Theta(\rho_1, \rho_2)$ is the Heaviside function which equals unity in the density region where the heater is active ($\rho_1 \leq \rho \leq \rho_2$) and zero elsewhere, and $\tau$ is the e-folding decay time-scale. The star reaches a quasi-stationary state which is regulated by the heat source \citep{Kaminker_2006, Kaminker_2007, Kaminker_2009, Kaminker_2014}. In Appendix \ref{sec:Heat_Test} we display the effect of the phenomenological heater on the internal temperature of the star for different values of $H_0, \rho_1$ and $\rho_2$. In line with previous results \citep{Kaminker_2006}, we find that it is easier for the source to heat the stellar surface when it is closer to the outer envelope, so that less of the thermal power supplied by the heater makes its way into the core, where it is dispersed by neutrino cooling. 

In this paper we determine how much thermal power generated by a phenomenological internal heat source $H$ is required by neutron stars with $B_{\textrm{dip}} \lesssim 4 \times 10^{13}$ G to match observed thermal luminosities $L_{\gamma} \gtrsim 10^{31}$ erg s$^{-1}$ at $t \gtrsim 10^{5.3}$ yr \citep{Potekhin_2020} of isolated neutron stars. By comparing the cooling curves calculated with an internal heater against $L_{\gamma}$ and age measurements, one can place constraints on the internal source.

\subsection{Comparing theory and observations}
\label{sec:theory_obs}
In Sections \ref{sec:Cooling_curves} and \ref{sec:Joule} we compare the theoretical cooling rates with the age and thermal luminosity measurements reported by \cite{Potekhin_2020}. Age and thermal luminosity measurements suffer from different sources of uncertainty, which we review briefly below.

\subsubsection{Age estimates}

The age of a source is often approximated by the characteristic age $\tau_c$, given by $\tau_c = P/(2\dot{P})$, where $P$ and $\dot{P}$ are the well-measured spin period and its time-derivative. However, $\tau_c$ provides a good estimate of the true age of the source only if $P$ at birth was much shorter than $P$ measured today and the dipolar magnetic field (responsible for the spin-down torque) has not decayed \citep{Vigano_2013}. These two conditions are not valid in general, and often $\tau_c$ overestimates the true age. In contrast, the \textit{kinematic age} does not depend on timing properties and assumptions about the rotational history, and it gives a more accurate estimate of the true age of the source. It can be inferred by supernova expansion models (assuming a distance), or by proper-motion measurements and an assumed birth sky position \citep{Cordes_1998, Hobbs_2005, Gaensler_2006, Motch_2009}. Only the very few cases of historical human records of supernovae whose remnants are associated to pulsars can pinpoint the exact age \citep{Green_2002, Green_2003}. 

In Section \ref{sec:Cooling_curves} we compare the theoretical cooling rates with the $L_{\gamma}$ and age measurements of isolated thermal emitters. If the age is given by $\tau_c$, it is considered to be an upper limit of the true age, and the data point is displayed in the plots with a leftward arrow. Otherwise, the age measurements and their error bars correspond to the kinematic age. 

\subsubsection{Thermal luminosity uncertainty}
Theoretical cooling curves can be compared either with the effective surface temperature inferred from observations (i.e. the temperature averaged over the stellar surface) or with thermal luminosity data. 

The effective surface temperature is inferred by fitting the observed spectra (which in general contain thermal and non-thermal components) with theoretical models. One systematic uncertainty is the emission model adopted to fit the thermal component of the spectrum (e.g. blackbody or light-element atmosphere). For some objects there is no clear indication that one emission model provides a better fit to the data with respect to alternative models (cf. \cite{Potekhin_2020} and references therein). Additionally, inhomogeneous surface temperatures caused by magnetic fields are realistically more complicated \citep{Igoshev_2021} than the limited number (usually a couple) of uniform-temperature regions of different sizes, commonly used in spectral modeling of observations (see e.g. \cite{Borghese_2021}). Moreover, in many cases the low photon counts and photoelectric absorption by the interstellar medium under $\sim$ 1 keV constitute further statistical and intrinsic uncertainties in $T_{\textrm{s}}$. 

Thermal luminosity measurements are conditional on the assumed spectral and absorption model, and on the composition of the envelope \citep{Yakovlev_2004, Vigano_2013, Potekhin_2014, Potekhin_2020}. Often, a dominant uncertainty in $L_{\gamma}$ originates in the poorly-constrained estimate of the distance of the source.

Given the above considerations, in this work we compare the theoretical cooling curves with the thermal luminosity data reported in \cite{Potekhin_2020}. In principle one should compare not only the theoretical and observed values of $L_{\gamma}$, but also the theoretical and observed surface maps of $T_\textrm{s}$, in order to account for the size of the emitting region on the stellar surface, for example.

\section{Cooling curves}
\label{sec:Cooling_curves}
In this section we present cooling curves of neutron stars with $B_{\textrm{dip}} \lesssim 4 \times 10^{13}$ G, with and without hyperon concentrations in the core, assuming different chemical compositions of the outer envelope. We look at the role of hyperon superfluidity in the absence of an internal heater (Section \ref{sec:Y_SF}). We then include a phenomenological heater in the stellar interior (Section \ref{sec:Internal_heat}), and quantify the parameters of the heater for stars with and without hyperon cores to match observations of magnetized, isolated thermally-emitting neutron stars with $B_{\textrm{dip}} \lesssim 4 \times 10^{13}$ G \citep{Potekhin_2020}. The simulations of stars with $B_{\textrm{dip}} \gtrsim 4 \times 10^{13}$ G and/or strong toroidal fields at birth are presented in Section \ref{sec:Joule}.

\begin{figure*}
\includegraphics[width=17cm, height = 12 cm]{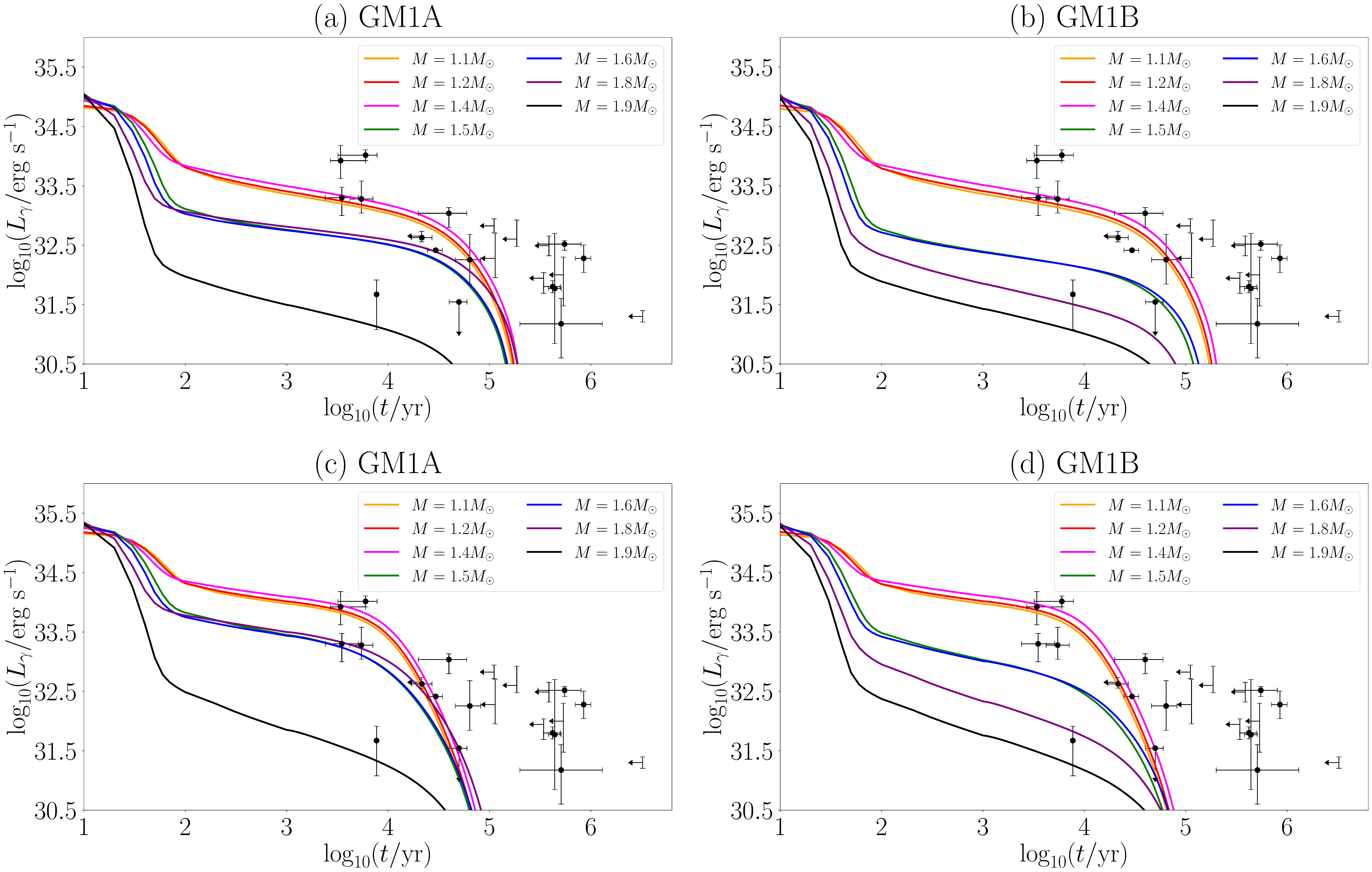}
\caption{Cooling curves of weakly magnetized neutron stars (with non-decaying, crust-confined poloidal dipole magnetic field with $B_{\textrm{dip}} = 10^{12}$ G) obtained with the GM1A and GM1'B EoSs. We consider stellar masses in the range $M \in [1.1, 1.9] \ M_{\odot}$. \textit{(a)} GM1A, iron-only envelope; nucleonic direct Urca is active for $M \gtrsim 1.1 \ M_{\odot}$. For $M \gtrsim 1.5 \ M_{\odot}$, hyperon direct Urca emission is also active. \textit{(b)} GM1'B, iron-only envelope. Nucleonic and hyperon direct Urca processes are active as for the GM1A EoS, but the proton and hyperon fractions are higher than the GM1A EoS, and the corresponding direct Urca emission is stronger. \textit{(c)} As in panel (a), but for a light-element envelope.  \textit{(d)} As in panel (b), but with a light-element envelope. Measurements of $L_\gamma$ and ages (points with error bars) for isolated stars are taken from \citet{Potekhin_2020}.}
\label{fig:M1}
\end{figure*}

\begin{figure*}
\includegraphics[width=17cm, height = 12cm]{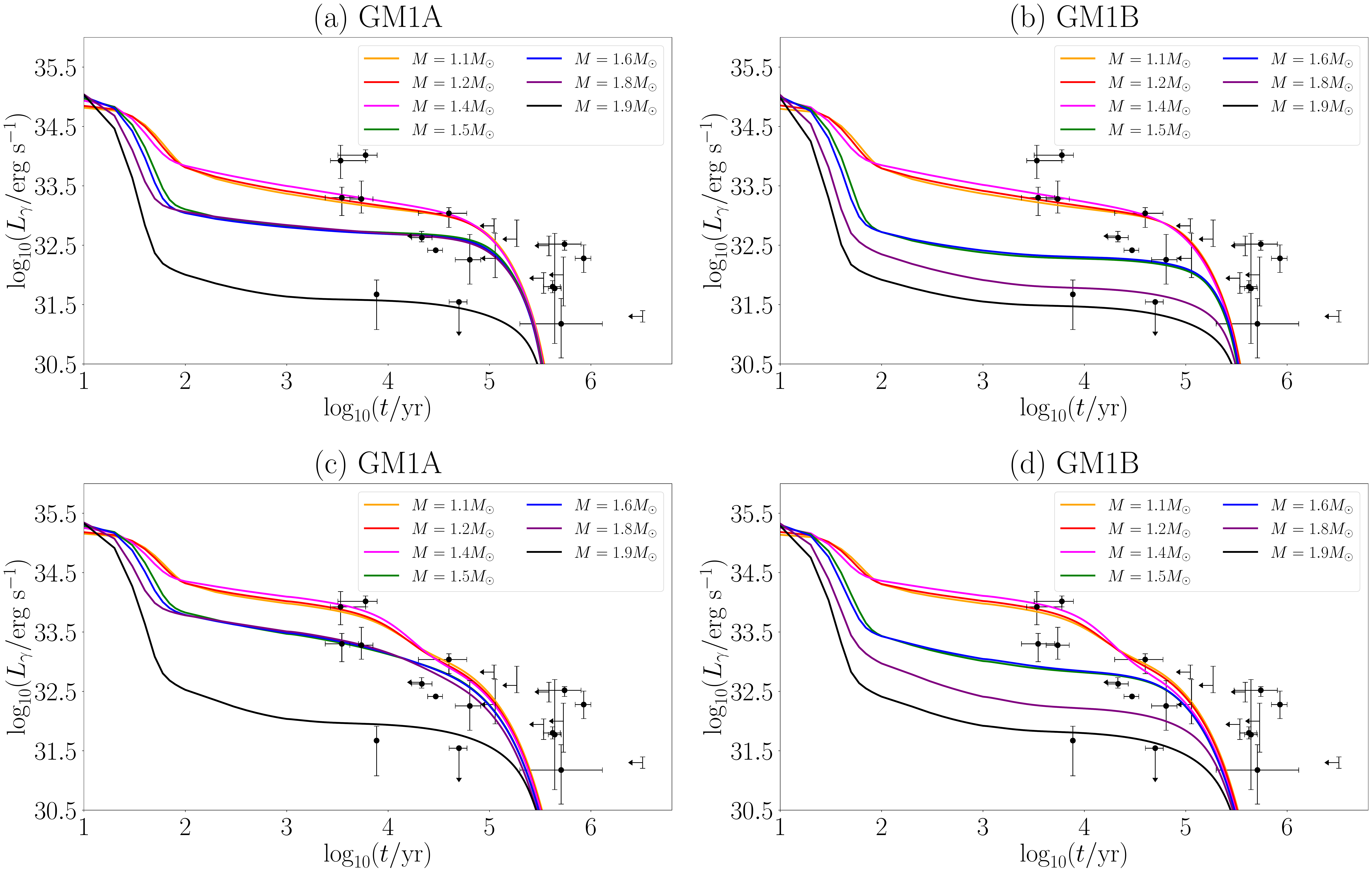}
\caption{Effect of a phenomenological heat source on the thermal evolution of weakly magnetized stars obtained with the GM1A EoS (panels (a) and (c)) and GM1'B EoS (panels (b) and (d)). The heat source is placed in the inner crust (in the layer with $10^{12} \ \textrm{g cm}^{-3} \leq \rho \leq 3.2 \times 10^{12} \ \textrm{g cm}^{-3}$) and produces a thermal power per unit volume $H_0 = 10^{16}$ erg cm$^{-3}$s$^{-1}$. \textit{(a)} GM1A, iron envelope. \textit{(b)} GM1'B, iron envelope. \textit{(c)} GM1A, light-element envelope. \textit{(d)} GM1'B, light-element envelope. All of the panels are calculated for a crust-confined, non-evolving poloidal (dipole) field with $B_{\textrm{dip}} = 10^{12}$ G. Data points taken from \citet{Potekhin_2020}.}
\label{fig:M2}
\end{figure*}

\begin{figure*}
\includegraphics[width=17cm, height = 12cm]{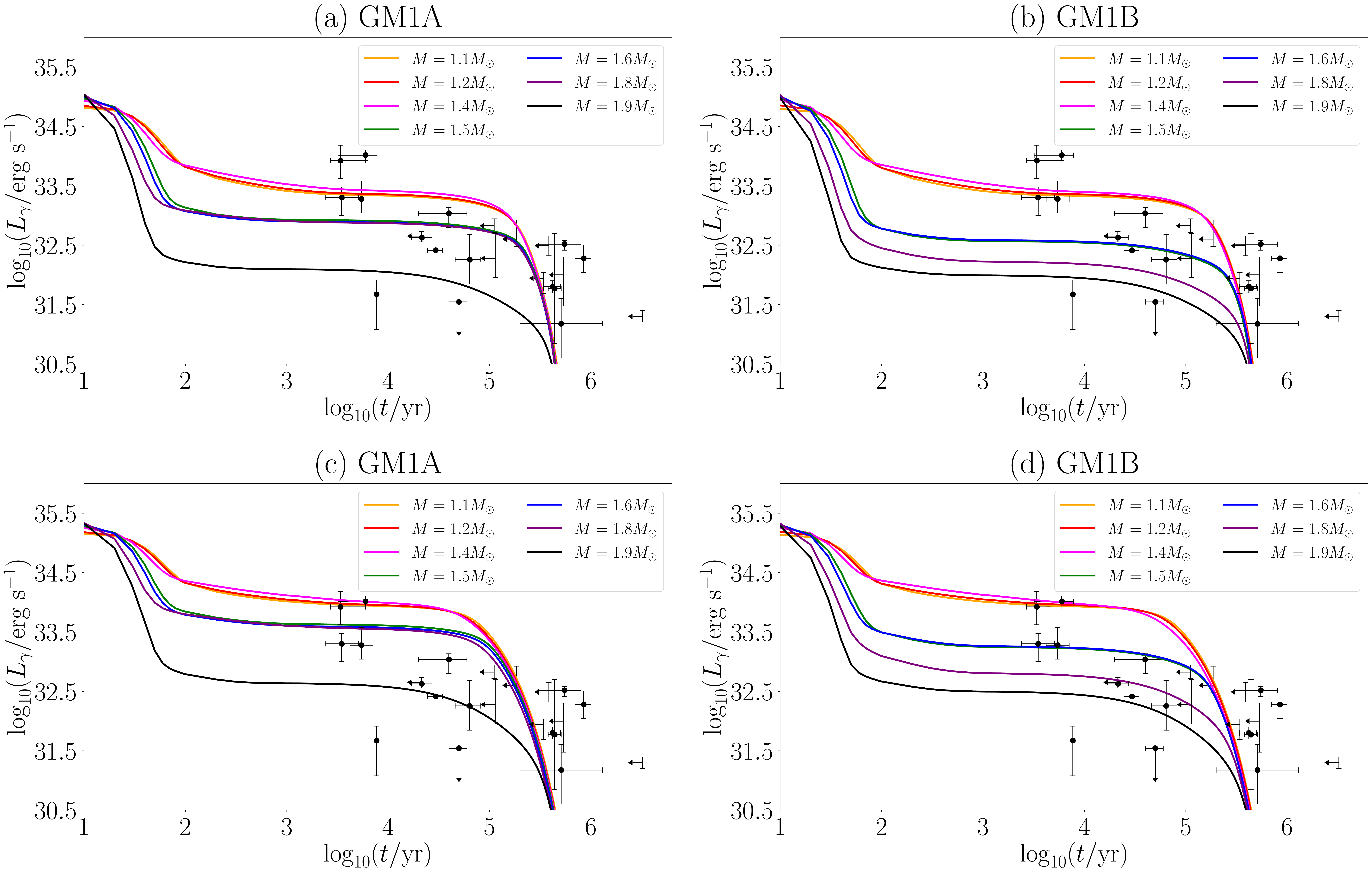}
\caption{As in Figure \ref{fig:M2}, but for $H_0 = 10^{17}$ erg cm$^{-3}$ s$^{-1}$.}
\label{fig:M3}
\end{figure*}

\subsection{Passive cooling}
\label{sec:Y_SF}
We calculate the thermal evolution of neutron stars without internal heating sources (passive thermal evolution) and with non-decaying, crust-confined weak poloidal dipolar magnetic fields (with surface value $B_{\textrm{dip}} = 10^{12}$ G at the pole and no toroidal component) as a cooling baseline and to validate the code against existing results in the literature. We note that since the magnetic field does not decay, the cooling models in this section do not include Joule heating. In the crust, the field causes non-radial heat transport. 

We simulate the thermal evolution of stars with masses in the range $1.1 \ M_{\odot} \leq M \leq 1.9 \ M_{\odot}$. Light masses in the range $1.2 \ M_{\odot} \lesssim M \lesssim 1.4 \ M_{\odot}$ are common across the neutron star population \citep{Ozel_2012}, while lower and higher masses are less common.   

In Figure \ref{fig:M1} we assume that nucleons and hyperons are superfluid. Hyperon superfluidity reduces the emissivity of the direct Urca processes involving hyperon species. Overall this causes the stars to cool down slower than when hyperons are not superfluid\footnote{Our code reproduces similar results to \cite{Raduta_2018} in the unmagnetized limit both when hyperons are in the normal phase and in the superfluid phase.} \citep{Raduta_2018}. Overplotted are some of the observed thermal luminosities and stellar age measurements reported in \citet{Potekhin_2020} of isolated thermally-emitting neutron stars (with $B_{\textrm{dip}} \lesssim 4 \times 10^{13}$ G and $L_{\gamma} \gtrsim 10^{31}$ erg s$^{-1}$ at $t \lesssim 10^{6.5}$ yr). Figures \ref{fig:M1}(a) and \ref{fig:M1}(b) display the thermal evolution of stars obtained with the GM1A and GM1'B EoSs respectively with a heavy-element envelope. In Figure \ref{fig:M1}(a), nucleonic direct Urca is active in a small fraction of the stellar volume in the model with $M = 1.1 \ M_{\odot}$ (orange curve). The region where nucleonic direct Urca is active increases in size for higher masses (e.g. for $M = 1.2 \ M_{\odot}$ and $M = 1.4 \ M_{\odot}$, red and magenta curves respectively). The curves match the thermal luminosities of some relatively young objects ($t \lesssim 10^5$ yr), such as CXOU J$085201.4$--$461753$ (with $L_{\gamma} \approx 2.0 \times 10^{33}$ erg s$^{-1}$ at $t \approx 3.5 \times 10^3$ yr) and PSR J$0538$+$2817$ (with $L_{\gamma} \approx 1.1 \times 10^{33}$ erg s$^{-1}$ at $t \approx 4.0 \times 10^4$ yr). At later stages of the thermal evolution, the cooling curves do not match the thermal luminosity of stars with $t \gtrsim 10^{5.3}$ yr, such as RX J$1605.3$+$3249$ (with $L_{\gamma} \approx 5.9 \times 10^{31}$ erg s$^{-1}$ at $t \approx 4.4 \times 10^5$ yr) or PSR J$0357$+$3205$ ($L_{\gamma} \approx 1.5 \times 10^{31}$ erg s$^{-1}$ at $t \approx 5.1 \times 10^{5}$ yr) for example.

Hyperons appear in heavier stars. The large neutron triplet and proton singlet gaps, combined with the $\Lambda$ and $\Xi^{-}$ pairing in the singlet channel, allow stars with $1.5 \ M_{\odot} \leq M \leq 1.8 \ M_{\odot}$ (i.e. with hyperon cores) to maintain relatively high thermal luminosities, with $L_{\gamma} \gtrsim 10^{32}$ erg s$^{-1}$ for $t \lesssim 10^5$ yr. Proton superfluidity combines with the pairing of $\Lambda$ hyperons, reducing the neutrino emissivity in the direct Urca channel involving $\Lambda$ and $p$ particles with respect to the case without hyperon superfluidity (cf. \cite{Raduta_2018}). Stars with masses in the range $1.5 \ M_{\odot} \leq M \leq 1.8 \ M_{\odot}$ overlap with the thermal luminosity and age measurements of relatively young objects such as PSR B$1951$+$32$ (with $L_{\gamma} \approx 1.8 \times 10^{32}$ erg s$^{-1}$ at $t \approx 6.4 \times 10^4$ yr) for example. For $M = 1.9 \ M_{\odot}$, both the $p$ and $\Lambda$ singlet gaps vanish close to the center of the star, the direct Urca emissivity involving $\Lambda$ and $p$ particles is not reduced by superfluidity of $\Lambda$ and $p$ particles, and the star cools down faster with respect to models with $M \lesssim 1.8 \ M_{\odot}$ \citep{Raduta_2018}. 

Figure \ref{fig:M1}(b) studies stellar models obtained with the GM1'B EoS. Neutron star models obtained with the GM1'B EoS have higher concentrations of $\Lambda$ and $p$ particles in the stellar core with respect to the GM1A EoS (for a fixed stellar mass). The thermal evolution of low-mass stars is similar to panel (a), but the cooling curves for $1.5 \ M_{\odot} \leq M \leq 1.8 \ M_{\odot}$ attain lower $L_{\gamma}$ with respect to panel (a) due to stronger direct Urca emission involving $\Lambda$ and $p$ particles.

We now consider light-element envelopes. We assume that the mass $\Delta M$ of light elements in the outer stellar envelopes satisfies $\Delta M / M \approx 10^{-7}$ \citep{Potekhin_1997, Potekhin_2003}. In general, stars with light-element envelopes have higher $L_{\gamma}$ than stars with iron-only envelopes in the neutrino dominated era. Models with light-element envelopes transition to the photon cooling stage earlier and do not match any of the data of mature stars with ages $t \gtrsim 10^5$ yr. As in Figure \ref{fig:M1}(a), the cooling curves in Figure \ref{fig:M1}(c) (GM1A EoS) are similar for $1.5 \ M_{\odot} \leq M \leq 1.8 \ M_{\odot}$, and lower values of $L_{\gamma}$ are attained by hyperon stars in Figure \ref{fig:M1}(b) (GM1'B EoS). Models with light-element envelopes are consistent with the thermal luminosity of young stars (with $L_{\gamma} \approx 10^{34}$ erg s$^{-1}$ at $t \lesssim 10^5$ yr) but are inconsistent with measurements corresponding to mature stars ($t \gtrsim 10^5$ yr).

In summary, stars with $M \leq 1.4 \ M_{\odot}$ and without internal heating are unable to explain the thermal luminosity of RX J$1856.5$--$3754$ (with $L_{\gamma} \approx 6.3 \times 10^{31}$ erg s$^{-1}$ at $t \approx 4.2 \times 10^5$ yr) or RX J$1605.3$+$3249$ for example.
The activation of hyperon direct Urca emission in massive stars ($M \gtrsim 1.4 \ M_{\odot}$) adds to nucleonic direct Urca processes to predict lower $L_\gamma$ than what is observed, and stars with hyperon cores are unable to explain the thermal luminosity of RX J$1605.3$+$3249$ or PSR J$0357$+$3205$ for example.
Although superfluidity of nuclear and hyperon species retards the cooling, Figure \ref{fig:M1} shows that the retardation is insufficient to explain the data if the neutrons pair in a large fraction of the stellar volume: stars with or without hyperon cores do not match the $L_{\gamma}$ measurements of objects with $t \gtrsim 10^{5.3}$ yr. To make matters worse, the thermal spectra of some mature X-ray isolated neutron stars (for example RX J$1605.3$+$3249$) can be fitted by hydrogen-atmosphere models \citep{Pires_2019, Potekhin_2020}. As shown in Figure \ref{fig:M1}, stars with light-element envelopes cool even faster than stars with iron-only envelopes, yielding $L_{\gamma} \lesssim 10^{30.5}$ erg s$^{-1}$ at $t = 10^5$ yr. 

In Appendix \ref{sec:Alternative_models} we consider different energy gap models for neutron triplet superfluidity, e.g. the \enquote{TToa} and \enquote{EEHOr} models \citep{Ho_2015}. Our conclusion do not change, if neutrons pair throughout the core and the energy gap has maximum amplitude satisfying $\gtrsim 0.4$ MeV (TToa model \citep{Ho_2015}). Two exceptions correspond to the scenario in which neutrons pair in the triplet channel only in part of the outer core, with energy gap $\lesssim 0.1$ MeV (EEHOr model \citep{Ho_2015}), or if they do not pair at all in the triplet channel. In this case, cooling models obtained with the GM1A EoS match most of the thermal luminosity and age data of isolated, thermally-emitting neutron stars with $B_{\textrm{dip}} \lesssim 4 \times 10^{13}$ G. 

We remind the reader that we do not account for the formation of Cooper pairs of nonidentical particles (nucleon-hyperon pairs). The latter may have interesting consequences in fast cooling high-mass stars for example, where the singlet gaps of protons and $\Lambda$ particles vanish close to the center of the star. If an attractive interaction exists among nonidentical particles whose Fermi surfaces are sufficiently close, cross-species superfluid phases may occur, cf. \cite{Sedrakian_2019} and references therein, as well as Figure 5 of \cite{Raduta_2018} for the particle density profiles corresponding to the GM1A EoS. This would suppress the emissivity of direct Urca processes in high-mass stars. However, the interaction potentials between nucleons and hyperons are hard to test experimentally, and it is difficult to assess the impact on the thermal evolution of neutron stars with hyperon cores.

\subsection{Phenomenological heating}
\label{sec:Internal_heat}

Although superfluid phases retard cooling, the observed thermal luminosities of neutron stars with $L_{\gamma} \gtrsim 10^{31}$ erg s$^{-1}$ and $B_{\textrm{dip}} \lesssim 4 \times 10^{13}$ G at $t \gtrsim 10^{5.3}$ yr are not attained by models with and $M \geq 1.1 \ M_{\odot}$ whether the envelope contains iron or light elements, if neutrons pair in a large fraction of the core. Here we employ the phenomenological model outlined in Section \ref{sec:Phenomenological_heat} to determine the effect of an internal heat source (of unspecified physical origin) on cooling. We use $\tau = 5 \times 10^4$ yr as in \cite{Kaminker_2006}, fix the region where the heater is active to $10^{12} \ \textrm{g cm}^{-3} \leq \rho \leq 3.2\times 10^{12}$ g cm$^{-3}$, and consider two values of $H_0$, namely $H_0 = 10^{16}$ erg cm$^{-3}$ s$^{-1}$ and $H_0 = 10^{17}$ erg cm$^{-3}$ s$^{-1}$. Other locations of the source and values of $H_0$ are studied in Appendix \ref{sec:Heat_Test}. The nucleon and hyperon superfluid gaps and the (non-evolving) magnetic field are the same as in Section \ref{sec:Y_SF}.

Figure \ref{fig:M2} displays the effect of a heat source (with $H_0 = 10^{16}$ erg cm$^{-3}$ s$^{-1}$, corresponding to $W(t) \approx 10^{33}$ erg s$^{-1}$ for $t \lesssim \tau$) on the cooling curves of stars obtained with the GM1A and GM1'B EoSs. The top panels in Figure \ref{fig:M2} refer to the GM1A EoS (panel (a)) and GM1'B EoS (panel (b)) for iron envelopes. The bottom panels study the same configurations as in the top panels, but for light-element envelopes. The additional thermal power supplied by the heater allows the cooling curves in Figure \ref{fig:M2}(a) to attain higher $L_{\gamma}$ than in Figure \ref{fig:M1}(a). Light stars ($M \leq 1.4 \ M_{\odot}$) without hyperons in the stellar core  match the thermal luminosity and age measurements of PSR J$0357$+$3205$ (contrarily to the case without internal heating, cf. Figure \ref{fig:M1}(a)) and cool down to $L_{\gamma} = 10^{30.5}$ erg s$^{-1}$ at later times than in Figure \ref{fig:M1}(a). Similar conclusions are valid for models hosting hyperons with $M = 1.5, 1.6, 1.8 \ M_{\odot}$, whose values of $L_{\gamma}$ are now slightly above the thermal luminosity of PSR B$0833$--$45$ (Vela pulsar; $L_{\gamma} \approx 4.2  \times 10^{32}$ erg s$^{-1}$ at $t \approx 2.1 \times 10^4$ yr). The effect of internal heating for models obtained with the GM1'B EoS and iron envelopes is studied in Figure \ref{fig:M2}(b), where the cooling curves of massive stars attain lower values of $L_{\gamma}$ than the GM1A EoS case due to the higher concentrations of proton and $\Lambda$ hyperons. Still, the cooling curves of stars with hyperon cores and $M = 1.5, 1.6, 1.8 \ M_{\odot}$ match thermal luminosity and age data of PSR B$1951$+$32$ and PSR J$0357$+$3205$, in contrast with Figure \ref{fig:M1}(b). 

The lower panels in Figure \ref{fig:M2} examine accreted, light-element envelopes.
Figure \ref{fig:M2}(c) studies models obtained with the GM1A EoS. The cooling curves fall below $L_{\gamma} = 10^{30.5}$ erg s$^{-1}$ at $t > 10^5$ yr (cf. Figure \ref{fig:M1}(c)), and agree with the measurements corresponding to stars with $L_{\gamma} \gtrsim 10^{32.5}$ erg s$^{-1}$ at $t \lesssim 10^5$ yr.
Figure \ref{fig:M2}(d) shows the light-element envelope case for the GM1'B EoS, where the cooling curves for stars with $M \gtrsim 1.4 M_{\odot}$ are lower than Figure \ref{fig:M2}(c).

We increase the intensity of the heat source up to $H_0 = 10^{17}$ erg cm$^{-3}$ s$^{-1}$ in Figure \ref{fig:M3}, leading to $W(t) \approx 10^{34}$ erg s$^{-1}$ for $t \lesssim \tau$. Generally speaking, this has the expected effect of raising the cooling curves and matching more of the data at $t \gtrsim 10^{5.3}$ yr. Figure \ref{fig:M3}(a) displays cooling of stellar models obtained with the GM1A EoS and with an iron-only envelope. The thermal power supplied by the heater is sufficiently high that the cooling curves of stars with or without hyperon cores overlap with RX J$1605.3$+$3249$ and PSR J$0357$+$3205$ for example.
Figure \ref{fig:M3}(b) considers a scenario similar to panel (a), but for the GM1'B EoS. Overall, the cooling curves of models with $M > 1.4 \ M_{\odot}$ attain lower $L_{\gamma}$ than the GM1A EoS case, but match the measurements of PSR B$1951$+$32$ and, at later times, of RX J$1605.3$+$3249$ and PSR J$0357$+$3205$. 

The agreement noted above is similar, when the iron-only envelope (top panels of Figure \ref{fig:M3}) is replaced by a light-element envelope (bottom panels of Figure \ref{fig:M3}).
In Figure \ref{fig:M3}(c) we study the same configuration as in Figure \ref{fig:M3}(a), but for accreted envelopes. The cooling curves show a better agreement with the observed thermal luminosities of objects whose thermal emission can be fitted with hydrogen-atmosphere models such as RX J$1605.3$+$3249$ \citep{Pires_2019, Potekhin_2020} for example, contrarily to the case without internal heating. Figures \ref{fig:M3}(d) displays the case of light-element envelopes for the GM1'B EoS. The cooling curves of massive hyperon stars attain lower $L_{\gamma}$ than panel (c).

In summary, our results show that a heat source located in the inner crust (in the density region $10^{12} \ \textrm{g cm}^{-3} \leq \rho \leq 3.2 \times 10^{12}$ g cm$^{-3}$) producing a thermal power per unit volume $H_0 = 10^{17}$ erg cm$^{-3}$ s$^{-1}$ (corresponding to $W(t) \approx 10^{34} \ \textrm{erg} \ \textrm{s}^{-1}$) allows magnetized models of low- and high-mass stars to agree better with optical and X-ray measurements of $L_{\gamma}$ of objects with $B_{\textrm{dip}} \lesssim 4 \times 10^{13}$ G. It is important to note that the inferred value of $W(t)$ depends, among other factors, on the superfluid model adopted, as noted in Section \ref{sec:superfluidity_th}. By changing the latter, the estimated $W(t)$ changes accordingly. Our main conclusions are similar to \cite{Kaminker_2006}, \cite{Kaminker_2007}, \cite{Kaminker_2009}. However there are some differences. First, \cite{Kaminker_2006}, \cite{Kaminker_2007}, \cite{Kaminker_2009} employ the phenomenological heater to match the models with observations of magnetars (with typical ages $t\lesssim 10^5$ yr and $L_{\gamma} \gtrsim 10^{34}$ erg s$^{-1}$), and infer $H_0 \gtrsim 10^{19}$ erg cm$^{-3}$ s$^{-1}$. We infer a lower $H_0$, which is sufficient for both low- and high-mass models to match observational data of isolated stars with $B_{\textrm{dip}} \lesssim 4 \times 10^{13}$ G at $t \gtrsim 10^{5}$ yr. Second, \cite{Kaminker_2006}, \cite{Kaminker_2007}, \cite{Kaminker_2009} include the effect of magnetic field only in the outer envelope. Our simulations account for the effect of the magnetic field in the crust, which causes non-radial heat transport. The latter effect is discussed in Appendix \ref{sec:Non-radial}.

\section{Joule heating}
\label{sec:Joule}

We now consider strongly magnetized stars. A complete study of the magneto-thermal evolution of stars with or without hyperon cores obtained with the GM1A and GM1'B EoSs lies outside the scope of this paper. Nevertheless, as a brief foretaste of what is possible, we present some preliminary results in this section for the GM1A EoS and two initial magnetic field configurations: (i) a mixed configuration with poloidal (dipole) and toroidal (quadrupole) components, with $B_{\textrm{dip}} = 10^{13}$ G and $\sim 90$ per cent of the magnetic stored in the toroidal component; and (ii) a purely poloidal dipolar magnetic field with $B_{\textrm{dip}} = 10^{14}$ G. In the following, we denote these two configurations A13T and A14 respectively. By way of illustration, we study the case of a $M = 1.8 \ M_{\odot}$ hyperon star with an iron-only envelope, assuming that nucleons and hyperons are superfluid as in Sections \ref{sec:Y_SF} and \ref{sec:Internal_heat}. 

\begin{figure}
\includegraphics[width=8.5cm, height = 7cm]{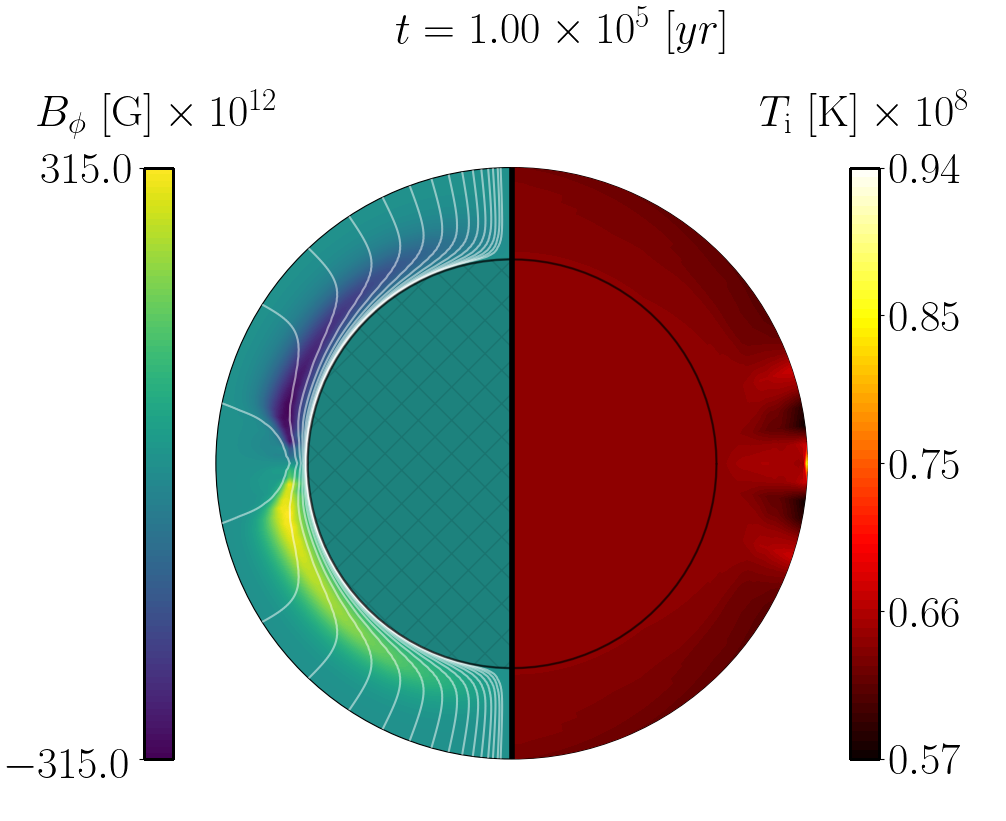}
\caption{Snapshot at $t = 10^{5}$ yr of the magneto-thermal evolution of a $M = 1.8 \ M_{\odot}$ star obtained with the GM1A EoS for the A14 configuration. In the left hemisphere, the contours show the toroidal field strength (left color bar; in units of $10^{12}$ G), and the white lines are the poloidal field lines. In the right hemisphere the contours show the redshifted internal temperature map (right color bar, in units of $10^8$ K). The crust (annulus) is enlarged for visualization purposes.}
\label{fig:POLAR}
\end{figure}

Figure \ref{fig:POLAR} shows the contours of the toroidal field $\boldsymbol{B}_{\textrm{tor}} = B_{\phi} \hat{\phi}$ and the poloidal field lines (left hemisphere), and the internal, redshifted temperature map ($T_{\textrm{i}} = T e^{\Phi}$, right hemisphere) at $t = 10^5$ yr for the A14 initial configuration. Hall drift distorts the magnetic field lines and generates a toroidal component whose maximum and minimum values of $B_{\phi}$ are located below the outer envelope for $t \lesssim 10^3$ yr. For $10^3 \ \textrm{yr} \lesssim t < 10^5$ yr, the maximum of the toroidal component moves towards the crust-core interface, where the resistivity is high due to pasta phases, causing further dissipation of the magnetic field \citep{Pons_2013}. Close to the equator, Hall dynamics generates small-scale magnetic structures where Ohmic dissipation is enhanced \citep{Pons_2019}. The temperature map at $t = 10^5$ yr shows that the core is isothermal (because of its high thermal conductivity), while the temperature distribution in the crust is anisotropic due to the difference in electronic transport across and along the magnetic field lines. In particular, the combination of Ohmic dissipation, anisotropic heat transport and the insulating effect of the tangential component of the magnetic field makes the equator region in the crust hotter than the polar region.

\begin{figure}
\includegraphics[width=8.5cm, height = 6.5cm]{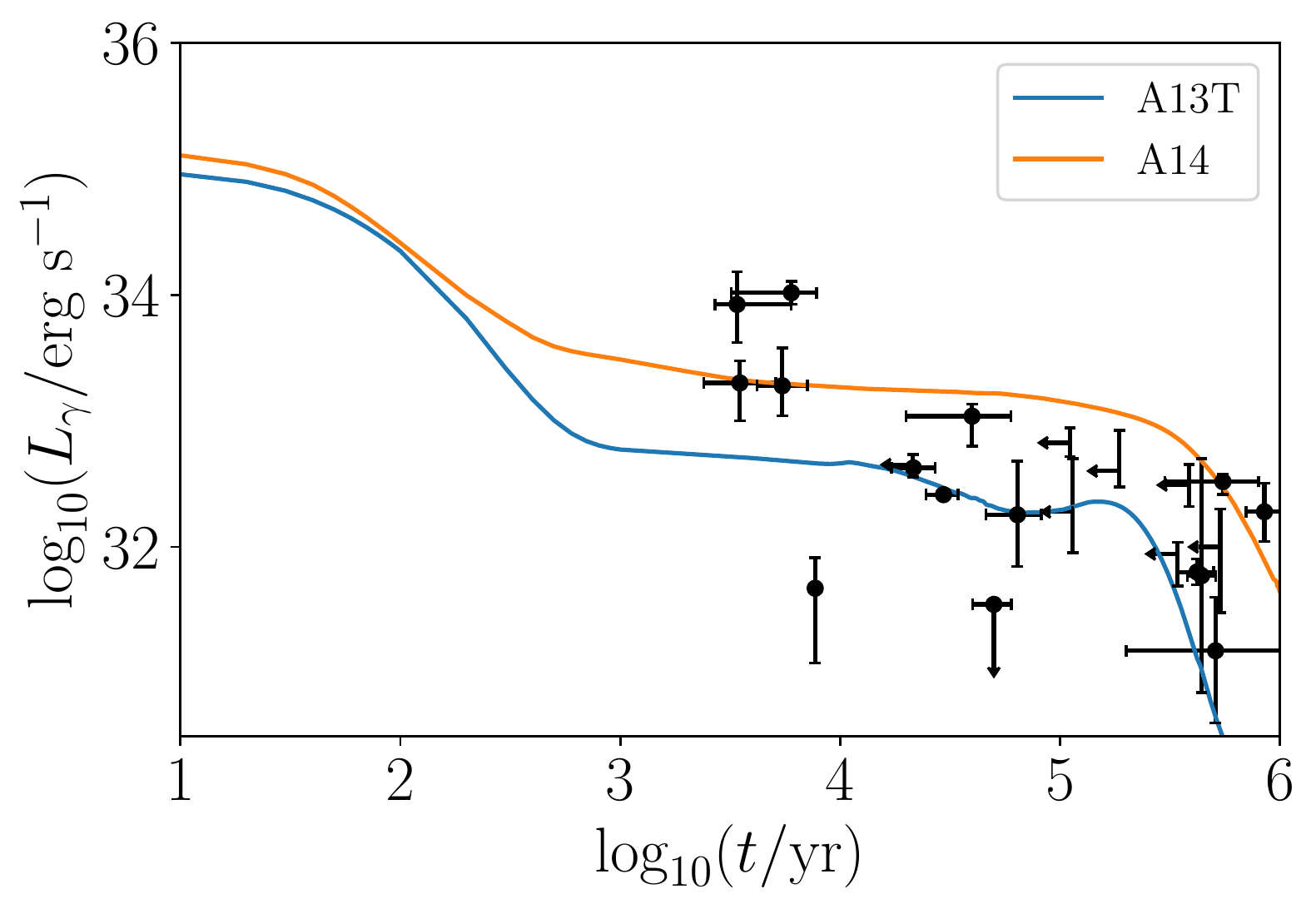}
\caption{Cooling curves including Joule heating for a $M = 1.8 \ M_{\odot}$ star (GM1A EoS, iron-only envelope) for the A13T (blue line) and A14 (orange line) initial magnetic configurations. Nucleon and hyperon species are superfluid. The data points are taken from \citet{Potekhin_2020}.}
\label{fig:2D_cool}
\end{figure}

The cooling curves for the A13T and A14 configurations are displayed in Figure \ref{fig:2D_cool}. They attain higher $L_{\gamma}$ than the case without internal heating (cf. Section \ref{sec:Y_SF}). The initial configuration A13T allows the cooling curve to overlap with data from Vela, RX J$1605.3$+$3249$ and PSR J$0357$+$3205$. The orange curve (A14 configuration) can explain even higher luminosities such as RX J$0720.4$--$3125$ ($L_{\gamma} \approx 1.9 \times 10^{32}$ erg s$^{-1}$ at $t \approx 8.5 \times 10^{5}$ yr).
We note also that for $t \gtrsim 10^5$ yr, the A13T and A14 configurations  attain $B_{\textrm{dip}} \approx 10^{12}$ G and $B_{\textrm{dip}} \approx 10^{13}$ G respectively. The latter values are compatible with the magnetic field strength inferred for the stars in the dataset employed in this work \citep{Potekhin_2020}.

In summary, Joule heating due to the decay of magnetic fields with $B_{\textrm{dip}} \gtrsim 10^{14}$ G or with strong toroidal fields at birth can explain $L_{\gamma} \gtrsim 10^{31}$ erg s$^{-1}$ at $t \gtrsim 10^{5.3}$ yr even for fast-cooling hyperon stars. Two-dimensional simulations are required, as the magnetic field affects the thermal conductivity of electrons, and the internal temperature in the crust becomes strongly inhomogeneous \citep{Vigano_2013, Pons_2019, Dehman_2020, Vigano_2021}. However, it is important not to overstate the result: it has been obtained for two arbitrary initial magnetic configurations. The comparison with the estimates for the phenomenological heater in Section \ref{sec:Internal_heat} is qualitative only, because the cooling curves in Figures \ref{fig:M2} and \ref{fig:M3} are calculated for weak fields ($B_{\textrm{dip}} \lesssim 10^{13}$ G), while Joule heating is important for $B_{\textrm{dip}} \gtrsim 10^{14}$ G and/or strong toroidal fields. A full-scale study of the magneto-thermal evolution of hyperon stars is left for future work.

\section{Conclusion}
\label{sec:end}

Fast cooling due to nucleonic and hyperon direct Urca can push $L_\gamma$ below its observed value ($10^{31} \lesssim L_{\gamma}/\textrm{erg s}^{-1} \lesssim 10^{33}$) for $t\gtrsim 10^{5.3}$ yr. In this paper we quantify the strength of a phenomenological heating source needed to restore agreement between the predicted and observed $L_\gamma$. We perform two-dimensional simulations of stars with and without hyperon cores and confirm that the thermal power inferred phenomenologically can indeed be provided by Joule heating starting from plausible poloidal and toroidal magnetic configurations.

Low-mass models ($1.1 \ M_{\odot}\lesssim M \lesssim 1.4 \ M_{\odot}$) and high-mass models ($1.5 \ M_{\odot}\lesssim M \lesssim 1.9 \ M_{\odot}$) obtained with the GM1A and GM1'B EoSs cool down fast due to direct Urca processes involving nucleon and hyperon species ($\Lambda$ and $\Xi^{-}$ particles). We find that if neutrons pair in the triplet channel in a large fraction of the stellar core (with maximum amplitude of the energy gap $\gtrsim 0.4$ MeV), stellar models with masses in the range $M \in [1.1, 1.9] \ M_{\odot}$ and no internal heat source do not match observations of isolated, thermally emitting neutron stars with $B_{\textrm{dip}} \lesssim 4 \times 10^{13}$ G and $L_{\gamma} \gtrsim 10^{31}$ erg s$^{-1}$ at $t \gtrsim 10^{5.3}$ yr \citep{Potekhin_2020}. 

We estimate the thermal power $W(t)$ required by weakly magnetized stars with or without hyperon cores and iron-only or light-element envelopes to explain observations of $L_{\gamma}$ \citep{Potekhin_2020} by invoking the presence of a phenomenological, internal heat source \citep{Kaminker_2006, Kaminker_2007, Kaminker_2009, Kaminker_2014}. For a typical heater lifetime of $\tau = 5\times 10^4$ yr, one finds that a heat source located in the inner crust supplying a thermal power per unit volume $H_0 =  10^{17}$ erg cm$^{-3}$ s$^{-1}$ for $t \lesssim \tau$ (corresponding to $W(t) \approx 10^{34}$ erg s$^{-1}$) allows models with $M \in [1.1, 1.9] \ M_{\odot}$ obtained with the GM1A and GM1'B EoSs to maintain $L_{\gamma} \gtrsim 10^{31}$ erg s$^{-1}$ for $t \gtrsim 10^{5.3}$ yr and to match the thermal luminosities of young (e.g. Vela pulsar) and mature (e.g. RX J$1605.3$+$3249$ and PSR J$0357$+$3205$) stars. Additionally, the thermal spectra of some sources (e.g. RX J$1605.3$+$3249$ \citep{Pires_2019, Potekhin_2020}) are compatible with hydrogen-atmosphere models. Internal heating allows cooling models with light-element envelopes to agree with observations. For $t \gtrsim \tau$, the initial heat power provided by the source decays exponentially, reaching $W(t) \approx 10^{33}$ erg s$^{-1}$ at $t = 10^5$ yr and $W(t) \lesssim 10^{30}$ erg s$^{-1}$ at $t = 5 \times 10^5$ yr. 

We emphasize that the $W(t)$ required to match observations depends on several factors. For example, it changes, if a different superfluid model is employed (e.g. considering different superfluid gaps for nucleons and hyperons and accounting for nucleon-hyperon pairing), and it is lower, if the age measurements employed in this work overestimate the true age \citep{Potekhin_2020}.

Joule heating due to the decay of magnetic fields with $B_{\textrm{dip}} \gtrsim 10^{14}$ G and/or strong toroidal fields is one plausible candidate for internal heating. A full study of Joule heating lies outside the scope of this paper, whose main goal is to quantify the strength and hence plausibility of a phenomenological heat source as an antidote to the fast cooling triggered by nucleonic and hyperon direct Urca. Nevertheless, by way of illustration, we examine the representative example of a $M = 1.8 \ M_{\odot}$ hyperon star obtained with the GM1A EoS and two initial magnetic field configurations: (i) an initial crust-confined, mixed poloidal (dipolar) and toroidal (quadrupolar) magnetic field configuration, with $B_{\textrm{dip}} = 10^{13}$ G and $\sim 90$ per cent of the total magnetic energy stored in the toroidal field; and (ii) an initial poloidal dipolar configuration with $B_{\textrm{dip}} = 10^{14}$ G ($B_{\phi} = 0$). Configuration (i) allows hyperon stars obtained with the GM1A EoS to match the measured $L_{\gamma}$ of Vela, RX J$1605.3$+$3249$ and PSR J$0357$+$3205$ for example. Configuration (ii) produces sufficient Joule heating rates to explain higher thermal luminosities ($L_{\gamma} \gtrsim 10^{32}$ erg s$^{-1}$), e.g. of RX J$0720.4$--$3125$.

It is important to remark that the results reported in this work are subject to a certain degree of arbitrariness related to the choice of the superfluid model. The latter affects the cooling rate of neutron stars and hence the required $W(t)$ to match observations. For example, following \cite{Raduta_2018} we consider only large hyperon energy gaps with maximum amplitude $\gtrsim 1$ MeV. Other works report smaller gaps for hyperon superfluidity \citep{Balberg_1998, Takatsuka_1999, Takatsuka_2001}. We also neglect cross-species pairing for simplicity, i.e. the formation of mixed nucleon-hyperon superfluidity, noting that several works find attractive interaction potentials among baryons of different species \citep{Zhou_2005, Nemura_2009, Meoto_2020, Haidenbauer_2020}.

The numerical calculations presented here follow in the footsteps of the extensive literature on neutron star cooling \citep{Yakovlev_1999, Yakovlev_2001, Page_2004, Page_2006, Vigano_2013, Potekhin_2015, Raduta_2018, Raduta_2019}. Other relevant physics, including other EoSs and different initial magnetic field configurations, are left for future work. We emphasize that in this work we focus primarily on the evolution of $L_{\gamma}$. More generally, though, one should study other observables too. For example, one should compare the theoretical and observed surface temperature maps. Unfortunately, this task is complicated by observational uncertainties and approximations often adopted when modeling the stellar surface from the observed spectra.

We conclude by emphasizing two caveats. First, the hypothesis of an internal heat source is hard to falsify, because many different physical heating mechanisms produce approximately the same $L_\gamma$. Conversely, given observations of $L_\gamma$, one can generally postulate more than one heat function $W(t)$ that matches the data. Hence one is not in a position to infer the nature of the heat source. Instead, one must be content with inferring its strength, if indeed it exists. Second, it is obvious that the tension between predictions and observations of $L_\gamma$ created by fast cooling due to hyperons goes away, if neutron stars do not contain hyperons. We do not seek to adjudicate the issue of whether hyperons exist in neutron stars or not, beyond noting that the theoretical arguments in their favor are plausible \citep{Glendenning_1985, Guichon_1988, Haensel_1994, Stone_2007}, but there are insufficient observational data at present to settle the matter one way or the other. The contribution of this paper is instead to extend existing studies \citep{Raduta_2018, Raduta_2019} to include internal heating, with more sophisticated multi-dimensional simulations to follow in future work.

\section*{Acknowledgements}
We thank the anonymous referee for valuable feedback and for drawing our attention to the issue of cross-species pairing. FA and AM thank Julian B. Carlin for useful discussions in the early stages of this work. FA is supported by the University of Melbourne through a Melbourne Research Scholarship. AM acknowledges funding from an Australian Research Council Discovery Project grant (DP170103625). DV is supported by the European Research Council (ERC) under the European Union’s Horizon 2020 research and innovation programme (ERC Starting Grant "IMAGINE" No. 948582, PI DV). CD is supported by the ERC Consolidator Grant “MAGNESIA” (No. 817661, PI Nanda Rea) and this work has been carried out within the framework of the doctoral program in Physics of the Universitat Aut\`onoma de Barcelona. JAP acknowledges support by the Generalitat Valenciana (PROMETEO/2019/071), AEI grant PGC2018-095984-B-I00 and the Alexander von Humboldt Stiftung through a Humboldt Research Award.

\section*{Data availability}
The numerical tables corresponding to the EoSs employed in this work are taken from the Web page \url{http://www.ioffe.ru/astro/NSG/heos/hyp.html}, and are calculated in \enquote{Physics input for modelling superfluid neutron stars with hyperon cores} \citep{Gusakov_2014}. 
The thermal luminosity and age data of neutron stars are reported in the paper \enquote{Thermal luminosities of cooling neutron stars} by \cite{Potekhin_2020} and are accessible at the Web page \url{http://www.ioffe.ru/astro/NSG/thermal/cooldat.html}.

\bibliographystyle{mnras}
\bibliography{BIBLIO}

\begin{appendix}

\section{Superfluid energy gaps}
\label{sec:nucleon_superfluidity}
\begin{table}
\caption{\label{tab:GAP_table} Parameters for the superfluid energy gaps for hyperon species in Eq. \eqref{eq:Y_gaps}.}
\begin{tabular}{cccccc}
\hline
Species&
\multicolumn{1}{c}{\textrm{$\Theta_0$ (MeV)}}&
\multicolumn{1}{c}{\textrm{$k_0$} (fm$^{-1}$)}&
\multicolumn{1}{c}{\textrm{$k_1$} (fm$^{-2}$)}&
\multicolumn{1}{c}{\textrm{$k_2$} (fm$^{-1}$)}&
\multicolumn{1}{c}{\textrm{$k_3$} (fm$^{-2}$)}\\
\hline

$\Lambda$ & \mbox{$18.5$} &\mbox{0.15} & \mbox{1.95}&\mbox{1.35} & \mbox{0.6}  \\
\textit{$\Xi^{-}$} & \mbox{$67.5$} &\mbox{0.1} & \mbox{3.2}&\mbox{2.7} & \mbox{11.0}  \\
\hline
\end{tabular}
\end{table}

In this appendix we describe the superfluid model adopted in this work.

Neutrons pair in the singlet channel ($^1\textrm{S}_0$ superfluidity) in the inner crust of the star, and in the triplet channel ($^3\textrm{P}_2$ -- $^3\textrm{F}_2$ superfluidity) in the core region. Protons, $\Lambda$ and $\Xi^-$ hyperons pair in the singlet channel in the core. We use the parameterizations reported in \cite{Ho_2015} for the energy gaps at zero temperature of neutrons (in the crust, \enquote{SFB} model) and protons (in the core, \enquote{CCDK}). The corresponding critical temperature is obtained using the relations in \cite{Ho_2015}. We follow \citet{Yanagi_2020} to determine the critical temperature for neutron triplet pairing (model \enquote{c} in \cite{Page_2004}, from \cite{Baldo_1998}). The critical temperature is given by

\begin{equation}
    T_{\rm{cn,t}} = T_{0}\exp\bigg[-\frac{(k_{\rm{F}}-k_0)^2}{(\Delta k)^2}\bigg] \ ,
\end{equation}
where $T_{0} = 10^{10}$ K, $k_{\rm{F}}$ is the Fermi wave number, $k_0 = 2.5$ fm$^{-1}$ and $\Delta k = 0.7$ fm$^{-1}$. We also consider different superfluid models for neutron triplet pairing, namely the \enquote{TToa} and \enquote{EEHOr} models \citep{Ho_2015}.

We calculate the singlet gap of $\Lambda$ and $\Xi^{-}$ hyperons at zero temperature adopting the parameterization in \cite{Kaminker_2002}, i.e.

\begin{eqnarray}
\Theta_{j, A}(k_{\rm{F}}) = \Theta_0 \frac{(k_{\rm{F}}-k_0)^2}{(k_{\rm{F}}-k_0)^2 + k_1}\frac{(k_{\rm{F}}-k_2)^2}{(k_{\rm{F}}-k_2)^2 + k_3} \ ,
\label{eq:Y_gaps}
\end{eqnarray}
where $j = \Lambda, \Xi^{-}$ and the parameters $\Theta_0$, $k_0$, $k_1$, $k_2$ and $k_3$ are given in Table \ref{tab:GAP_table}. The corresponding gaps are similar to the ones calculated in \cite{Raduta_2018} for the GM1A EoS.

The temperature-dependent energy gaps are calculated using the formulae reported in \cite{Levenfish_1994} and \cite{Yakovlev_1999}.
For singlet pairing, the temperature dependent gap is given by

\begin{equation}
    \frac{\Theta_{l, A}(T)}{k_\textrm{B} T} = \sqrt{1-\frac{T}{T_{\rm{c}}}}\bigg(1.456 - \frac{0.157}{\sqrt{T/T_{\rm{c}}}} + \frac{1.764}{T/T_{\rm{c}}}\bigg) \ ,
    \label{eq:singlet_gap}
\end{equation}
where $l = n, p, \Lambda, \Xi^{-}$, $k_\textrm{B}$ is the Boltzmann constant and $T_{\rm{c}}$ denotes the corresponding critical temperature for simplicity. For triplet pairing (with projected total momentum of the pair $m_J = 0$), the neutron superfluid gap can be fitted by

\begin{equation}
    \frac{\Theta_{n, B}(T)}{k_\textrm{B} T} = \sqrt{1-\frac{T}{T_{\rm{c}}}}\bigg(0.7893 + \frac{1.188}{T/T_{\rm{c}}} \bigg)\sqrt{1 + 3 \cos^2\gamma} \ .
    \label{eq:triplet_gap}
\end{equation}
The gap $\Theta_{n,B}$ is anisotropic, as it depends on the angle $\gamma$ between the momentum vector of the particle and the quantization axis. All of the quantities required for the cooling simulations that depend on the angle $\gamma$ are integrated over the solid angle and are given in \cite{Yakovlev_1999}.

\section{Alternative superfluidity models}
\label{sec:Alternative_models}

\begin{figure*}
\includegraphics[width=17cm, height = 12 cm]{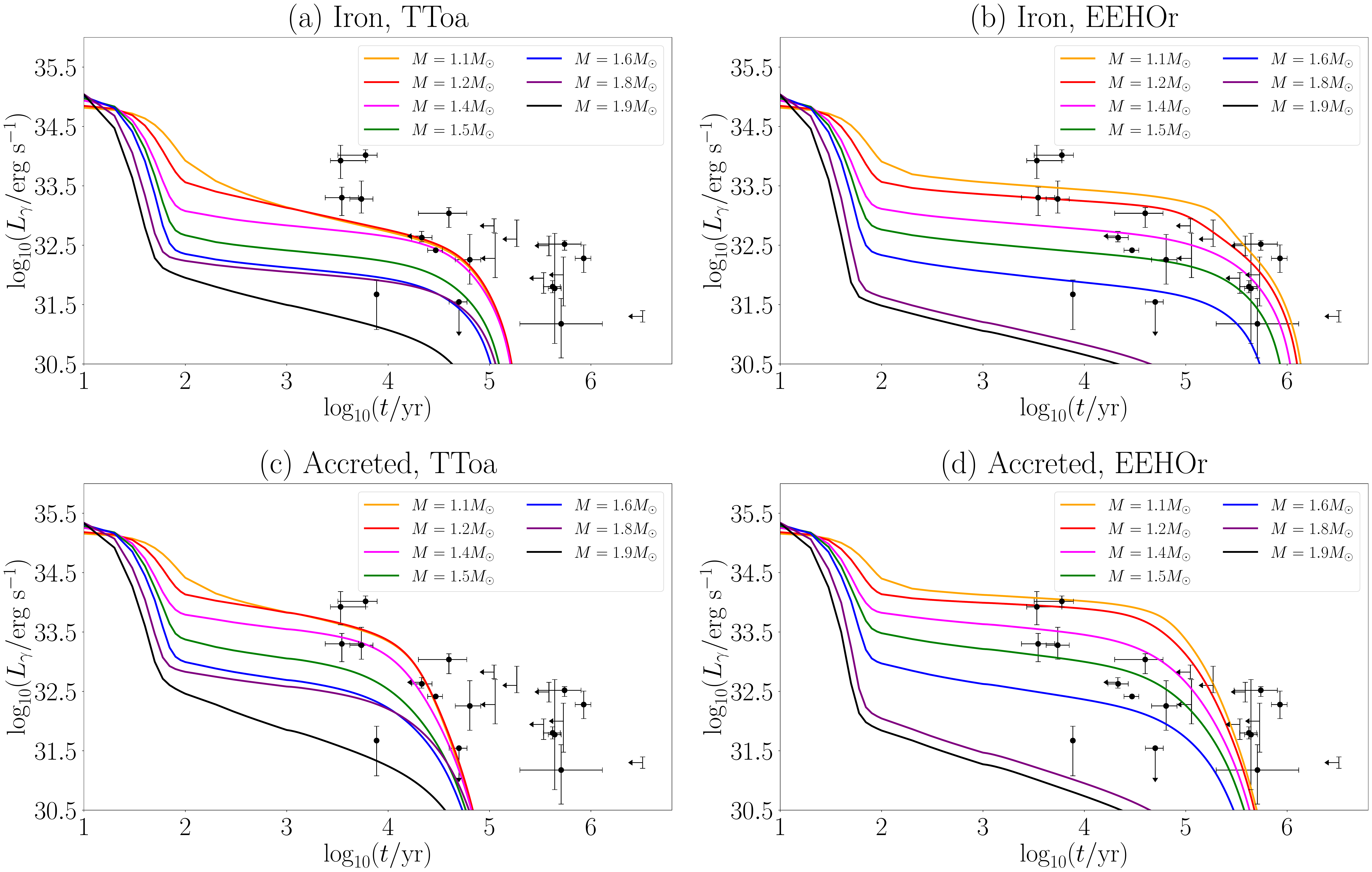}
\caption{Cooling curves of models obtained with the GM1A EoS calculated using the \enquote{TToa} and \enquote{EEHOr} models for neutron triplet superfluidity \citep{Ho_2015}. \textit{(a)} Iron envelope, TToa model. \textit{(b)} Iron envelope, EEHOr model. \textit{(c)} As in panel (a), but for a light-element envelope. \textit{(d)} As in panel (b), but with a light-element envelope. The magnetic field configuration is the same as in Figure \ref{fig:M1}.}
\label{fig:Other_Gaps}
\end{figure*}

In this appendix, we study the cooling of stars obtained with the GM1A EoS with iron-only and accreted envelopes, assuming different superfluid models for neutron triplet pairing (\enquote{TToa} and \enquote{EEHOr} models \citep{Ho_2015}). The energy gaps of protons and hyperons and the non-decaying magnetic field are the same as in Section \ref{sec:Cooling_curves}. In panels (a) and (b) of Figure \ref{fig:Other_Gaps} we consider iron-only envelopes (TToa and EEHOr models respectively), and in panels (c) and (d) we consider accreted envelopes (TToa and EEHOr models respectively). The critical temperature for neutron triplet superfluidity in the TToa model (with $T_{\textrm{cn,t}} \lesssim 6\times 10^8$ K) is lower than model \enquote{c} \citep{Page_2004} used in Section \ref{sec:Cooling_curves}, and neutrons pair throughout the stellar core. For the EEOHr model, one has $T_{\textrm{cn,t}} \lesssim 2\times 10^8$ K, and neutrons are unpaired in a large fraction of the core.

In Figure \ref{fig:Other_Gaps}(a), the cooling curves calculated with the TToa model are in general lower with respect to Figure \ref{fig:M1}(a). Models with $M \leq 1.4 M_{\odot}$ overlap with the measurements of PSR B$0833$--$45$ (Vela pulsar) and PSR B$1951$+$32$, and do not overlap with the measurements of stars with ages $t \gtrsim 10^{5.3}$ yr.
For $M \gtrsim 1.5 M_{\odot}$, the volume where nucleonic direct Urca emission is active increases in size, and the presence of $\Lambda$ and $\Xi^{-}$ hyperons triggers direct Urca processes involving hyperon species. Although the superfluidity of nucleons and hyperons reduces the emissivity of the direct Urca processes, the critical temperature for neutron triplet pairing is lower with respect to the one employed in Section \ref{sec:Cooling_curves}, the energy gap is smaller and the corresponding nucleonic direct Urca emission stronger, leading to faster cooling rates.  

Figure \ref{fig:Other_Gaps}(b) shows cooling of stars obtained with the GM1A EoS and iron envelopes, but with the EEHOr neutron triplet gap. The main difference with Section \ref{sec:Cooling_curves} and Figure \ref{fig:Other_Gaps}(a) is that the emissivity of nucleonic direct Urca processes is reduced only by the large proton gap while neutrons are unpaired in most of the core. However, the neutron heat capacity is not reduced by neutron triplet superfluidity, retarding noticeably the cooling. Low-mass stars with $M = 1.1 \ M_{\odot}$ are compatible with RX J$1308.6+2127$ (with $L_{\gamma} \approx 3.3 \times 10^{32}$ erg s$^{-1}$ at $t \approx 5.5 \times 10^5$ yr), RX J$1605.3$+$3249$ and PSR J$0357$+$3205$. By increasing the mass, the cooling curves overlap with CXOU J$085201.4$--$461753$, PSR B$1951$+$32$, RX J$1605.3$+$3249$ and PSR J$0357$+$3205$.

We study the case of accreted envelopes in Figures \ref{fig:Other_Gaps}(c) and \ref{fig:Other_Gaps}(d) for the TToa and EEHOr gaps respectively. Similarly to the results in Figure \ref{fig:M1}(c), the cooling curves in Figure \ref{fig:Other_Gaps}(c) do not match the data of stars with ages $t \gtrsim 10^5$ yr. Different results are found in Figure \ref{fig:Other_Gaps}(d). The model with $M = 1.1 \ M_{\odot}$ overlaps with CXOU J$185238.6$+$004020$ ($L_{\gamma} \approx 1.0 \times 10^{34}$ erg s$^{-1}$ at $t \approx 6.0 \times 10^3$ yr), and is compatible at later times with data corresponding to RX J$1605.3$+$3249$. Higher mass stars overlap with the measurements of PSR J$1119$--$6127$ (with $L_{\gamma} \approx 1.9 \times 10^{33}$ erg s$^{-1}$ at $t \approx 5.5 \times 10^3$ yr), PSR B$1951$+$32$ and PSR J$0357$+$3205$ for example.

We conclude that, for the TToa model, neutron stars require internal heating to match observations of objects with ages $t \gtrsim 10^{5.3}$ yr. Instead, if the gap is small (EEHOr model), neutron stars' models obtained with the GM1A EoS do not require internal heating to match $L_{\gamma}$ observations.

\section{Heat source in the inner crust}
\label{sec:Heat_Test}

\begin{figure*}
\includegraphics[width=17cm, height = 6.3 cm]{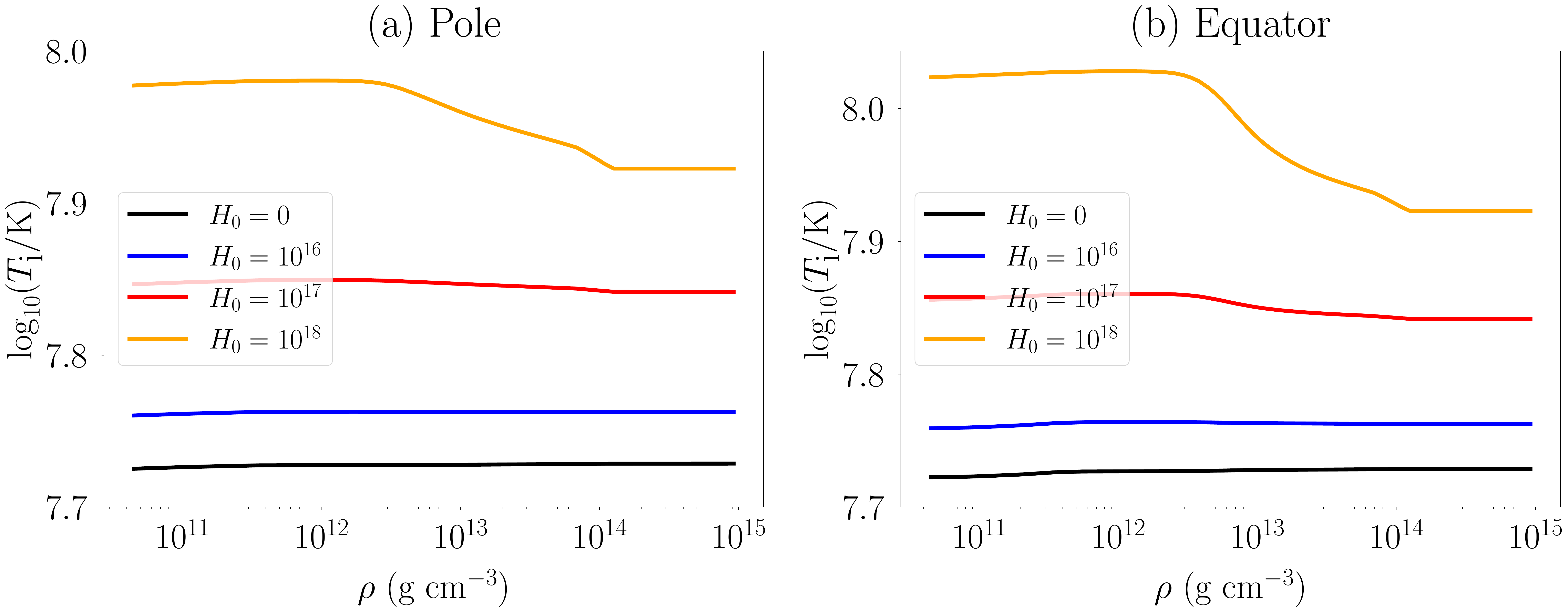}
\caption{Internal redshifted temperature $T_\textrm{i}$ at $t = 10^4$ yr versus density in a star with mass $M = 1.8 \ M_{\odot}$ (GM1A EoS) for different values of the thermal power per unit volume ($H_0$ in units of erg cm$^{-3}$ s$^{-1}$) generated by a phenomenological heat source. The magnetic field is a non-decaying, crust-confined poloidal dipole with $B_{\textrm{dip}} = 10^{12}$ G.}
\label{fig:Heat_Source_2D}
\end{figure*}

\begin{figure*}
\includegraphics[width=17cm, height = 6.3 cm]{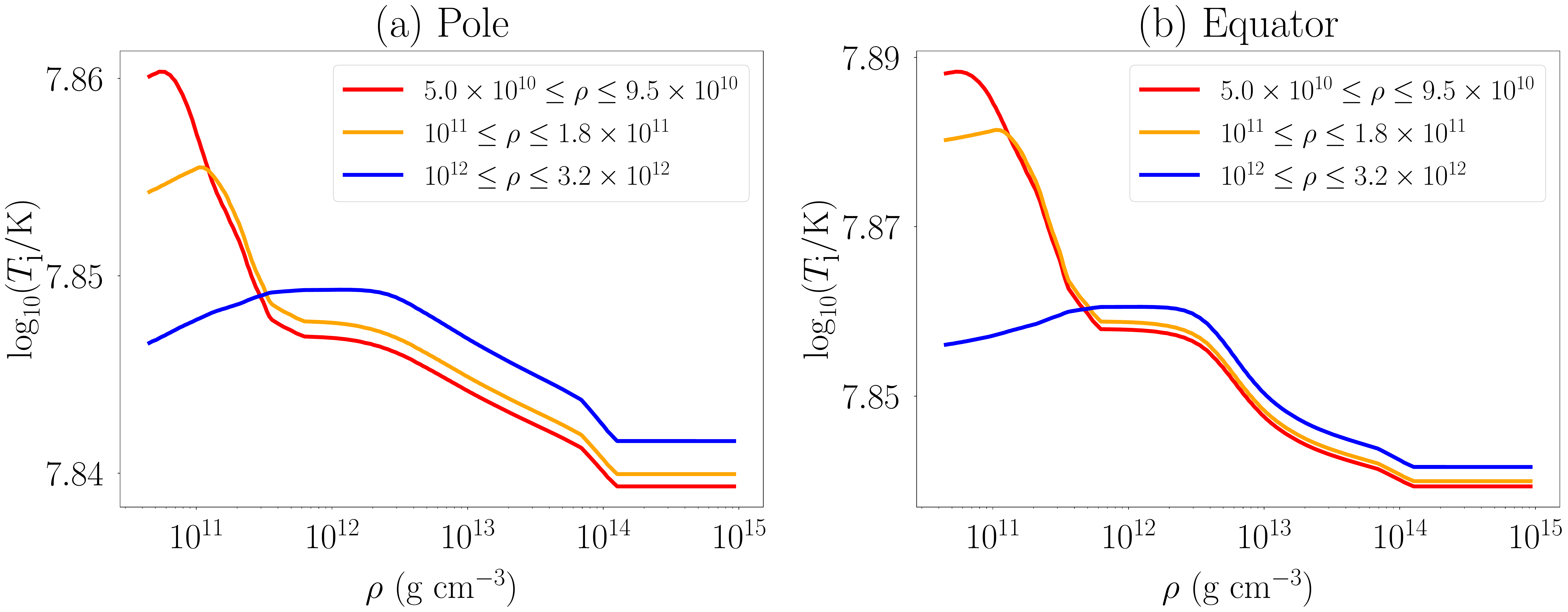}
\caption{As for Figure \ref{fig:Heat_Source_2D}, but with $H_0 = 10^{17}$ erg cm$^{-3}$ s$^{-1}$ fixed and the internal heat source confined to three layers of different thickness (colored curves; see legend for corresponding density range in units of g cm$^{-3}$).}
\label{fig:Heat_Source_POSITION}
\end{figure*}

In this appendix we check how the depth where the internal heater is located and $H_0$ affect the internal, redshifted temperature of the star.

In Figure \ref{fig:Heat_Source_2D}, we fix the density interval in which the heat source operates to $10^{12} \ \textrm{g} \ \textrm{cm}^{-3} \leq \rho \leq 3.2 \times 10^{12}$ g cm$^{-3}$ and we use $\tau = 5 \times 10^4$ yr (as in \cite{Kaminker_2006}). As a concrete example, we consider a neutron star with $M = 1.8 \ M_{\odot}$ obtained with the GM1A EoS with an iron-only envelope, for a crust-confined, non-evolving poloidal (dipole) magnetic field with $B_{\textrm{dip}} = 10^{12}$ G (as in Section \ref{sec:Cooling_curves}). We show the internal redshifted temperature $T_\textrm{i} = T e^{\Phi}$ at $t = 10^4$ yr at the pole and at the equator. In general the temperature difference $\Delta T_{\textrm{bc}}$ between the bottom of the outer envelope ($\rho_\textrm{b} \approx 4 \times 10^{10}$ g cm$^{-3}$ at the base) and the core increases with $H_0$. For $H_0 = 10^{18}$ erg cm$^{-3}$ s$^{-1}$ we find $\Delta T_{\textrm{bc}} / T_{\textrm{b}} \approx 12$ per cent at the pole and $\approx 21$ per cent at the equator, where $T_{\textrm{b}}$ denotes the temperature at the bottom of the outer envelope. For $H_0 = 10^{17}$ erg cm$^{-3}$ s$^{-1}$, we find $ \Delta T_{\textrm{bc}} / T_{\textrm{b}} \approx 1$ per cent at the pole and $\Delta T_{\textrm{bc}} / T_{\textrm{b}} \approx 3$ per cent at the equator. 

The depth at which the heat is deposited affects the temperature at the boundary of the isothermal region due to the thermal conductivity of the crust layers \citep{Kaminker_2006}. For example, in Figure \ref{fig:Heat_Source_POSITION} we consider the same model as in Figure \ref{fig:Heat_Source_2D} ($H_0 = 10^{17}$ erg cm$^{-3}$ s$^{-1}$) but vary the density interval where the heat is deposited. The closer the source lies to the outer envelope ($\rho_\textrm{b} \approx 4 \times 10^{10}$ g cm$^{-3}$), the higher is the corresponding $\Delta T_{\textrm{bc}}/T_{\textrm{b}}$. For the red line, the maximum difference is $\Delta T_{\textrm{bc}}/T_{\textrm{b}}\approx 5$ per cent at the pole and $\Delta T_{\textrm{bc}}/T_{\textrm{b}} \approx 11$ per cent at the equator, which is larger than the values of $\Delta T_{\textrm{bc}}/T_{\textrm{b}}$ due to the presence of a heat source deeper in the inner crust (blue line) discussed above.

In conclusion, the source maintains high $T_\textrm{i}$ at the bottom of the outer envelope more easily, if it lies closer to the outer envelope. For the purposes of this work, a source located in the inner crust layer with $10^{12} \ \textrm{g cm}^{-3} \leq \rho \leq 3.2 \times 10^{12} \ \textrm{g cm}^{-3}$ (i.e. not adjacent to the outer envelope) is sufficient for low-mass and high-mass stars obtained with the GM1A and GM1'B EoSs to explain observations of isolated thermally-emitting neutron stars with $B_{\textrm{dip}} \lesssim 4\times 10^{13}$ G and $L_{\gamma} \gtrsim 10^{31}$ erg s$^{-1}$ at $t \gtrsim 10^{5.3}$ yr as shown in Section \ref{sec:Internal_heat}.

\section{Non-radial heat transport with an internal heater}
\label{sec:Non-radial}

\begin{figure*}
\includegraphics[width=14cm, height = 7cm]{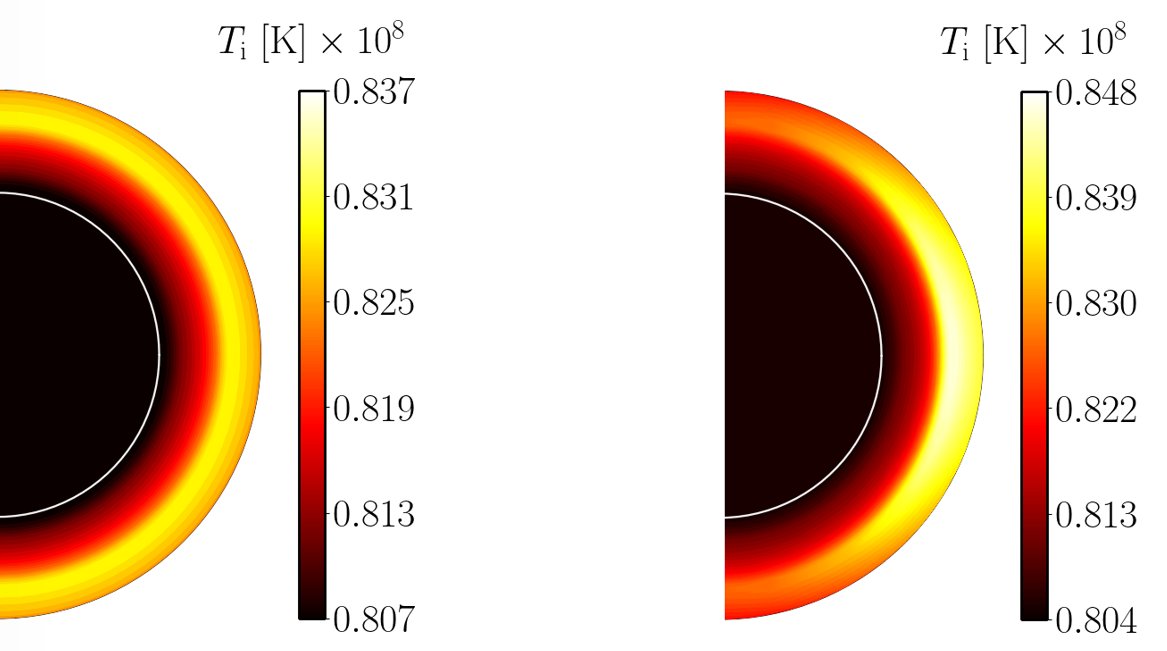}
\caption{Maps of the redshifted internal temperature $T_{\textrm{i}}$ in the presence of a phenomenological heat source in an unmagnetized star (left panel) and in a magnetized star (right panel) with a crust-confined, poloidal dipolar and non-evolving field with $B_{\textrm{dip}} = 10^{12}$ G. The mass of the star is $M = 1.5 \ M_{\odot}$ (GM1A EoS). The heater produces a thermal power per unit volume $H_0 = 10^{17}$ erg cm$^{-3}$ s$^{-1}$ and is located in the density layer with $10^{12} \ \textrm{g cm}^{-3} \leq \rho \leq 3.2 \times 10^{12}$ g cm$^{-3}$. The superfluid gaps are the same as in Figure \ref{fig:M3}, and the snapshots are taken at $t = 10^4$ yr. The crust (annulus) is enlarged for visualization purposes, and the solid white line marks the crust-core interface.}
\label{fig:Heat_Map}
\end{figure*}

As discussed at the end of Section \ref{sec:Internal_heat}, the presence of a magnetic field causes non-radial heat transport. The latter effect is accentuated in the presence of an internal heat source. 

By way of illustration, we show in Figure \ref{fig:Heat_Map} a snapshot of the map of the redshifted internal temperature $T_{\textrm{i}} = T e^{\Phi}$ at $t = 10^4$ yr for a $M = 1.5 \ M_{\odot}$ star obtained with the GM1A EoS and with an iron-only envelope. As in Section \ref{sec:Internal_heat}, we place a phenomenological heat source in the region with density $10^{12} \ \textrm{g cm}^{-3} \leq \rho \leq 3.2 \times 10^{12}$ g cm$^{-3}$, which produces a thermal power per unit volume $H_0 = 10^{17}$ erg cm$^{-3}$ s$^{-1}$. The left map displays the unmagnetized case. The right map displays the magnetized case with a crust-confined, non-evolving dipolar poloidal field with surface strength $B_{\textrm{dip}} = 10^{12}$ G at the pole. In the left panel in Figure \ref{fig:Heat_Map}, there is a brighter, thin annulus in the crust where $T_{\textrm{i}}$ is higher than the surrounding regions due to the thermal power supplied by the heater, reaching $T_{\textrm{i}} \approx 8.3 \times 10^7$ K. The right panel in Figure \ref{fig:Heat_Map} shows the effect of non-radial heat transport caused by the magnetic field. The bright annulus is not visible, as heat is transported by the electrons along the magnetic field lines towards the equator, where it is stored due to the insulating effect of the tangential field. In the region where the heater is active, we have $T_{\textrm{i}} \approx 8.3 \times 10^7$ K at the poles, and $T_{\textrm{i}} \approx 8.5 \times 10^7$ K at the equator. The value of $T_{\textrm{i}}$ beneath the outer envelope (the latter is not shown in the plots) is higher in the right map (magnetized case) with respect to the left map (unmagnetized case). As a result, the thermal luminosity for the magnetized model is $L_{\gamma} \approx 7.6 \times 10^{32}$ erg s$^{-1}$, while in the unmagnetized model $L_{\gamma} \approx 5.7 \times 10^{32}$ erg s$^{-1}$.

\end{appendix}





\bsp	
\label{lastpage}
\end{document}